\documentclass[11pt]{article}

\usepackage{amsmath, amsfonts, amssymb, amsthm}
\usepackage{graphicx}

\usepackage{times}
\usepackage{bm}
\usepackage[authoryear]{natbib}

\newcommand{\RR}{\mathbb{R}}
\newcommand{\NN}{\mathbb{N}}
\newcommand{\PP}{\mathrm{pr}}
\newcommand{\EE}{E}
\newcommand{\C}{\mathcal{C}}

\newcommand{\sd}{\,\mathrm{d}}
\newcommand{\Es}{E}
\renewcommand{\Pr}{\mathrm{pr}}

\newcommand{\Var}{\mathrm{var}}
\newcommand{\LS}{\mathrm{LS}}
\newcommand{\cens}{\mathrm{cens}}

\newtheorem{theorem}{Theorem}

\newtheorem{example}{\normalfont \scshape Example}
\newtheorem{remark}{\normalfont \scshape Remark}

\begin{document}

\title{Extremal Behavior of Aggregated Data \\
  with an Application to Downscaling}

\author{Sebastian Engelke\thanks{\'Ecole Polytechnique F\'ed\'erale de Lausanne, EPFL-FSB-MATHAA-STAT, 
       Station 8, 1015 Lausanne, Switzerland. Email: sebastian.engelke@epfl.ch}, \;
       Rapha\"el de Fondeville \thanks{\'Ecole Polytechnique F\'ed\'erale de Lausanne, EPFL-FSB-MATHAA-STAT, 
       Station 8, 1015 Lausanne, Switzerland. Email: raphael.de-fondeville@epfl.ch} \; and \;
       Marco Oesting \thanks{Universit\"at Siegen, Department Mathematik, Walter-Flex-Str.~3, 57068 Siegen, Germany.
       Email: oesting@mathematik.uni-siegen.de}}
\maketitle

\begin{abstract}
The distribution of spatially aggregated data from a stochastic process $X$ may exhibit a different tail behavior than its marginal distributions.
  For a large class of aggregating functionals~$\ell$ we introduce the $\ell$-extremal coefficient 
  that quantifies this difference as a function of the extremal spatial dependence
  in~$X$.
   We also obtain the joint extremal dependence for multiple aggregation functionals applied to the same process.
    Explicit formulas for the $\ell$-extremal coefficients and multivariate
  dependence structures are derived in important special cases.
  The results provide a theoretical link between the extremal distribution of the aggregated data
  and the corresponding underlying process, which we exploit to develop a
  method for statistical downscaling. We apply our framework to downscale daily temperature maxima
  in the south of France from a gridded data set and use our model to generate high resolution maps of the warmest day during the $2003$ heatwave.
\end{abstract}


\section{Introduction}

Spatial extreme value theory and, especially, 
max-stable processes are widely applied tools to assess risks in environmental science. 
These processes are motivated by the study of
\begin{equation}\label{eq:Mn}
  M_n(s) =   \max_{i=1,\ldots,n} \frac{X_i(s) - b_s(n)}{a_s(n)}, \quad s \in S, 
\end{equation}
where $X_1,\dots, X_n$ are independent observations of a sample-continuous
process $X$, modeling a phenomenon of interest such as rainfall or temperature on some region $S$. 
The scaling functions $a_s(n) > 0$ and $b_s(n) \in \RR$, $n \in \NN$, are both continuous in $s \in S$. 
Functional limits obtained from this construction
as $n\to \infty$, named the class of max-stable processes, are appealing models
for spatial extremes. Their realizations, however, are 
composed of different single events $X_i$, which prohibits direct interpretation 
and renders efficient inference and simulation challenging \citep[e.g.,][]{dom2016a, thi2015}.

It is often more natural to study threshold exceedances, or, more precisely, the 
extremal behavior of $\ell(X_i)$, $i=1,\dots,n$, where $\ell$ is a functional on
the space of continuous functions on~$S$. \cite{bui2008}, for instance, 
consider the daily rainfall over a certain region~$S$, and therefore choose 
$\ell(X) = \int_S X(s) \mathrm d s$. Using the same functional, \cite{col1996} 
relate the tail of the distribution of the integral to the tail of the distribution 
at a single location, and \cite{FDHZ12} formalize this idea through the so-called reduction factor. For general homogeneous functionals 
$\ell$, \cite{dom2016} characterize the functional limits of threshold exceedances 
$u^{-1} X \mid \ell(X) > u$, for a high threshold $u$.

In this paper we follow the approach of \cite{col1996} and \citet{FDHZ12} in order 
to investigate the tail behavior of more general functionals $\ell$. Under 
certain conditions we show that, for any $s_0 \in S$,
\begin{equation} 
   \PP\left[\frac{\ell(X) - \ell\{b_s(n)\}}{\ell\{a_s(n)\}} > x\right] \approx \theta^\ell \PP\left\{\frac{X(s_0) - b_{s_0}(n)}{a_{s_0}(n)} > x\right\} , \quad x \in \RR,
\end{equation}
for sufficiently large $n$. This means that the tail of the $\ell$-functional 
of $X$ behaves like the tail at an individual location times a reduction factor
$\theta^\ell$, which we call the $\ell$-extremal coefficient. In different contexts,
the interpretation of this coefficient might differ, but intuitively $\theta^\ell$
summarizes the effect of spatial extremal dependence in $X$ on the risk diversification through the functional~$\ell$. 

The $\ell$-extremal coefficient relates the tail of the univariate random variable
$\ell(X)$ to the multivariate or spatial extremal dependence in $X$. A major
advantage of this functional perspective is that it produces return level estimates
that are consistent with respect to the underlying structure of $X$, even when 
considering different aggregation functionals applied to the same process $X$. 
Indeed, for functionals $\ell_1,\dots, \ell_L$, we study the multivariate tail
behavior of $(\ell_1(X), \dots, \ell_L(X))$, which turns out to be in the 
max-domain of attraction of a multivariate max-stable distribution. 

Popular models for the functional limit of the maxima $M_n$ in \eqref{eq:Mn} are Brown--Resnick processes that take a similar
role in spatial extremes as Gaussian processes in classical 
geostatistics. The reason for this is that the former are essentially the only
such limits when $X$ is a stationary Gaussian
process and an additional rescaling is allowed \citep{kab2009}. This connection can be exploited to perform efficient 
inference \citep{wadsworth-tawn14, eng2014, thi2015} and simulation
\citep{DEMR13, dom2016a, oes2017} for Brown--Resnick processes based on densities
and sampling algorithms of Gaussian random vectors. In our framework, this link
to Gaussian distributions allows us to use results from the geostatistical literature
on data aggregation \citep[e.g.,][]{Wackernagel2003} to obtain explicit expressions
for $\theta^\ell$ and the extremal dependence in $(\ell_1(X), \dots, \ell_L(X))$ if the limiting process $Z$ 
in \eqref{eq:Mn} is a Brown--Resnick process with Gumbel margins.

An important consequence of our findings is that they allow, under certain
assumptions, to recover the tail distribution of $X$ based only on information
from the aggregated vector. This is similar to inferring the extremal dependence
of $X$ based only on extremal coefficients \citep[cf.,][]{sch2003}.
In meteorology, for instance, large scale climate models provide 
only data over grid cells, but practical questions require risk assessment
at point locations such as cities or other infrastructural sites. 
Techniques to perform this transition from large to small scales are summarized under the
notion of downscaling. 
In the second part of the paper we thus propose a statistical downscaling method to
infer in a spatially consistent way the tail behavior of the underlying stochastic process $X$
based on the observed extremes of the aggregated data. Relevant outputs
will be the exceedance probabilities at point locations and simulations 
of spatial extreme events of $X$, both unconditionally and conditionally on the
observed aggregated extremes.
We apply this procedure to coarse scale gridded temperature data in the south of France from the e-obs data set
\citep{Haylock2008}. The fitted model provides, for instance, fine-resolution simulations of the warmest day during the $2003$ heatwave conditionally on the observed grid values.

\section{Limit results for extremes of aggregated data} \label{sec:theory}

\subsection{Background on extremes}

Let $S$ be a compact subset of a complete separable metric space. By $C(S)$ we
denote the space of real-valued functions on $S$ equipped with the supremum
norm $\|\cdot\|_\infty$, defined by $\|f\|_\infty = \sup_{s \in S} |f(s)|$, and
the corresponding Borel $\sigma$-algebra $\C(S)$. 

We consider a sample-continuous stochastic process $\{X(s), \, s \in S\}$, which
we assume to be in the max-domain of attraction of a max-stable process with
common marginal extreme value index $\xi \in \RR$. More precisely, 
for independent copies $X_1,\ldots,X_n$ of $X$,
there exist functions $a_s: (0,\infty) \to (0,\infty)$, $b_s: (0,\infty) \to \RR$, both 
continuous in $s\in S$, such that as $n \to \infty$, the process $M_n$ of componentwise
maxima defined in \eqref{eq:Mn} converges in distribution on the space $C(S)$, i.e.,
\begin{equation} \label{eq:x-conv}
 \mathcal{L}\left\{ \max_{i=1,\ldots,n} \frac{X_i(s) - b_s(n)}{a_s(n)}, \, s \in S\right\} \longrightarrow
 \begin{cases}
   \mathcal{L}\{ \text{sgn}(\xi)Z(s)^{\xi},  \, s \in S\}, & \xi\neq0,\\
   \mathcal{L} \{\log Z(s),     \, s \in S\}, & \xi=0,\\
 \end{cases}
\end{equation}
where $\mathcal{L}(\eta)$ denotes the law of a process $\eta$. By definition, the process $Z$ in the limit is max-stable, and it is simple in the sense
that it is normalized to have unit Fr\'echet margins \citep[cf.,][Chapter 9]{DHF06}. Moreover,
for any $s\in S$, the margin $X(s)$ then is in the max-domain of attraction of an extreme 
value distribution
\begin{equation}
 \label{gev}G_{\xi}(x) = 
 \begin{cases} 
   \exp\left[-\{\text{sgn}(\xi)x\}^{-1/\xi}\right],   & \quad \xi\neq 0, \\
   \exp\left\{-\exp(-x)\right\},        & \quad \xi = 0,
 \end{cases}
\end{equation}
for all $x \xi \geq 0$. The different distributions are called $(1/\xi)$-Fr\'echet 
for $\xi>0$, Gumbel for $\xi=0$ and $(-1/\xi)$-Weibull for $\xi<0$, respectively.
The assumption of a spatially constant $\xi$ in \eqref{eq:x-conv} is common in the
literature since it is required to obtain meaningful theoretical results,
and it is usually a reasonable hypothesis in applications.

According to its spectral representation \citep[cf.][]{dehaan84, GHV90, penrose92},
\begin{equation} \label{eq:spec-repr}
Z(s) = \max_{i \in \NN} U_i W_i(s),  \quad s \in S, 
\end{equation}
where $\{U_i, i \in \NN\}$ are the points of a Poisson point process on
$(0,\infty)$ with intensity measure $u^{-2} \sd u$ and the spectral functions $W_i$, $i \in \NN$, are
independent copies of some non negative sample-continuous process
$\{W(s), \, s \in S\}$ with $\EE \{W(s)\} = 1$ for all $s \in S$. 

In the sequel we assume that $X$ is non negative for the Fr\'echet case $\xi > 0$, 
while for the Weibull case $\xi < 0$ each $X(s)$, $s \in S$, is assumed to have the same upper endpoint $0$.
Finally, we provide the example of the widely used class of Brown--Resnick max-stable processes.
\begin{example}\label{ex:BR}
  Let $\{G(s): s\in S\}$ be a centered Gaussian
  process with variogram $\gamma(s,t) = \Var\{G(s)-G(t)\}$.
  A Brown--Resnick process is the max-stable process $Z$ in \eqref{eq:spec-repr} where 
  the spectral functions follow the distribution of 
  $$W(s) = \exp\left[G(s) - \Var\{G(s)\} / 2 \right], \quad s \in S. $$
  The distribution of $Z$ only depends on the variogram $\gamma$, and
  for $s_1,\dots, s_m\in S$, the finite dimensional distribution of 
  $(Z(s_1), \dots, Z(s_m))$ is called the H\"usler--Reiss distribution
  \citep{hue1989} with parameter matrix $\Gamma = \{\gamma(s_j,s_k)\}_{j,k=1,\dots,m}$;
  more details can be found in Appendix \ref{sec:HR} 
  and in \cite{bro1977}, \cite{kab2009} and \cite{kab2011}. 
\end{example}

\subsection{Univariate limiting distributions of aggregated data}
We first derive the univariate asymptotic distribution of aggregated data.
Following \citet{FDHZ12}, we assume that the normalizing functions $a_s(t)$ 
can be decomposed asymptotically into positive functions $A(s)$ and $a(t)$
in the sense that
\begin{equation} \label{eq:decomp-a}
 \sup_{s \in S} \left|\frac{a_s(t)}{a(t)} - A(s)\right| \to 0, \quad \text{as } t \to \infty.
\end{equation}
In the Gumbel case $\xi=0$, the left-hand side of \eqref{eq:decomp-a} is 
even assumed to be equal to zero, i.e.,
\begin{equation} \label{eq:decomp-a-gumbel}
  a_s(t) = A(s) a(t) \quad \text{for all } s \in S, \ t>0. 
\end{equation}
For data aggregation, we consider a positively homogeneous functional
$\ell: C(S) \to \RR$, i.e., $\ell$ satisfies $\ell(af) = a \ell(f)$ for all $a > 0$, $f \in C(S)$.
We further assume that $\ell$ is uniformly continuous, and we use the notation
$\ell(f)$ and $\ell\{f(s)\}$ interchangeably.

The following theorem is a particular case of Theorem \ref{thm2}, that is proved 
in Section \ref{subsec:multiv-asymptotics}. Alternatively it can be proved similarly
to \citet[Theorem 2.1]{FDHZ12}.

 \begin{theorem}\label{thm1}
 Let $\ell$ be a positively homogeneous and uniformly continuous functional on
 $C(S)$. Further, assume that \eqref{eq:x-conv} and \eqref{eq:decomp-a} hold.
 If $\xi \leq 0$, the spectral functions $W$ belonging to the process $Z$
 in \eqref{eq:x-conv} are assumed to be strictly positive. Then, for 
 $\xi \neq 0$, we have
 \begin{equation} \label{eq:conv-ell-univ}
  \lim_{t \to \infty} t \: \PP\left[\frac{\ell(X)}{ a(t)\ell\{A(s)\}} > x\right] = \theta_\xi^\ell |x|^{-1/\xi}, \quad \xi x > 0,
 \end{equation}
 where
 \begin{equation} \label{eq:theta}
   \theta_\xi^\ell = \EE\left(\left[\frac{\ell\{W(s)^\xi A(s)\}}{\ell\{A(s)\}}\right]^{1/\xi}\right).
 \end{equation}
 For $\xi =0$, we further require that \eqref{eq:decomp-a-gumbel} holds and
 that $\ell$ is linear. In this case, 
 \begin{equation} \label{eq:conv-ell-gumbel-univ}
   \lim_{t \to \infty} t \: \PP\left[\frac{\ell(X) - \ell\{b_s(t)\}}{a(t)\ell\{A(s)\}} > x\right] = \theta_0^\ell \exp\left(-{x}\right), \quad x \in \RR,
 \end{equation}
 where
 \begin{equation} \label{eq:theta-0}
   \theta_0^\ell = \EE\left\{\exp\left(\frac{\ell\{\log[W(s)] A(s)\}}{\ell\{A(s)\}}\right)\right\}.
 \end{equation}
\end{theorem}
\begin{remark}
      Theorem \ref{thm1} is formulated for threshold exceedances, but, using well-known 
      equivalences from univariate extreme value theory, it could be easily reformulated
      to describe the limiting behavior of $\max_{i=1}^n \ell(X_i)$, where $X_1,\dots, X_n$ 
      are independent copies of $X$.
\end{remark}

We call the quantity $\theta_\xi^\ell$ the $\ell$-extremal coefficient since it
describes the change of the upper tail of the $\ell$-aggregated data compared
to the tail of the univariate marginal data. Our definition of $\theta_\xi^\ell$
in Theorem~\ref{thm1} contains a normalization by $\ell\{A(s)\}$, making it 
invariant under multiplication of $\ell$ by a constant and thus simplifying 
interpretation. Indeed, for $\xi>0$ and $s_0\in S$, we observe that
$$ \theta_\xi^\ell = \lim_{u\to \infty} \frac{\PP\left\{\ell(X)/ \ell(A) > u\right\}}{\PP\left\{X(s_0)/ A(s_0) > u\right\}}.$$
The interpretation of this coefficient might differ depending on the respective
context and risk functional $\ell$. In general, $\theta_\xi^\ell$ 
summarizes the effect of the spatial extremal dependence in $X$ on the 
diversification of the risk through functional $\ell$. Importantly, not only
the dependence but also the marginal tail index $\xi$ may effect the 
coefficient $\theta_\xi^\ell$, which we stress in Theorem~\ref{thm1} and 
henceforth by the index $\xi$. 

The concept of the $\ell$-extremal coefficient extends and unifies various 
notions in extreme value statistics and applied sciences such as extremal
coefficients, diversification factors in portfolios and areal reduction factors.
We present these and other examples for illustration, always assuming that
$X$ satisfies the conditions of Theorem \ref{thm1}.

\begin{example} \label{ex:int}
  The important case where $S\subset \RR^2$ is a compact region and 
  $\ell(f) = 1/|S|\int_S f(s) \, \mathrm{d}s$ was first studied in 
  \citet{col1996} and \citet{bui2008} in the framework of total areal rainfall, and 
  then formalized by \citet{FDHZ12}. In this case of a spatial average, the coefficient 
  $\theta_\xi^\ell = \theta_\xi^{\text{avg}}$ is popular in environmental science 
  where it is called the areal reduction factor. Hydrologists use it to convert 
  quantiles of point rainfall to quantiles of total rainfall over a river catchment
  of interest. Interestingly, this coefficient satisfies $0<\theta_\xi^{\text{avg}}\leq 1$
  for $\xi \leq 1$, and $\theta_\xi^{\text{avg}}\geq 1$ for $\xi \geq 1$ \citep[][Prop.~2.2]{FDHZ12}.
  That means that average rainfall is less extreme than point rainfall if the marginals
  have finite expectation, as typically encountered in practice, and more extreme if they have infinite expectation.
\end{example}

\begin{example}\label{ex:port}
  If $S = \{s_1,\dots, s_m\}$ is a finite set and $\ell(f) = \sum_{i=1}^m c_i f(s_i)$ 
  is a weighted sum with fixed $c_1,\dots, c_m \geq 0$, then \citet{zhou} and 
  \citet{mainik_embrechts} computed the corresponding coefficient 
  $\theta_\xi^\ell$ for $\xi>0$. In this setup, $X(s_i)$, $i=1,\dots,m$,
  are interpreted as dependent, heavy-tailed risk factors, and $\theta_\xi^\ell$
  represents the diversification in the portfolio $P = \sum_{i=1}^m c_iX(s_i)$.
  More precisely, the value at risk of $P$ for high levels $\alpha\to 1$ can be
  expressed as the value at risk of a single factor times a constant that involves
  the $\ell$-extremal coefficient $\theta_\xi^\ell$. Theorem \ref{thm1} yields an analogous 
  result also for light tailed risk factors.
\end{example}

\begin{example}
  Another well-known example is the case of $S=\{s_1,\ldots,s_m\}$ being a finite
  set and $\ell(f) = \max_{i=1}^m f(s_i)$. If we further have $A(s_1)=\ldots=A(s_m)$, 
  then $\theta_\xi^\ell = \EE\left\{\max_{i=1}^m W(s_i)\right\}$ corresponds to
  the classical extremal coefficient \citep{sch2003}, a number 
  between $1$ and $m$ that is usually interpreted as the number of asymptotically 
  independent random variables among $X(s_1),\dots,X(s_m)$. A similar interpretation is valid also if $S$ is an 
  arbitrary compact subset, and $\theta_\xi^\ell = \EE\left\{\max_{s \in S} W(s)\right\}$
  is a spatial extension of the classical extremal coefficient.
\end{example}

\begin{example}
  As a last example, we consider energy functionals of the type 
  $\ell^2(f) = \int_S f(s)^2 \, \mathrm{d}s$ that appear in various applications
  in physics. In the case of $X$ being a wind field, $\ell^2(X)$ is proportional
  to the integrated kinetic energy over a region $S$, which is an indicator for 
  the potential damage caused by the corresponding storm event \citep[eg.][]{powell07}. 
\end{example}

The expressions \eqref{eq:theta} and \eqref{eq:theta-0} for the $\ell$-extremal
coefficient are expected values of functions of the spectral process $W$. The 
distribution of the latter is known for most popular models, and it includes 
truncated Gaussian processes \citep{sch2002,opi2013} and $\log$-Gaussian
processes \citep{bro1977, kab2009}, for instance. Numerical evaluation of 
$\theta_\xi^\ell$ is thus readily implemented through simulations of $W$. In the 
important case of $\xi=0$ and $W$ corresponding to a log-Gaussian process, we 
obtain a closed form expression for $\theta_0^{\text{avg}}$.

\begin{example}\label{ex1}
Suppose that $\xi=0$ and $Z$ is a Brown--Resnick process on a compact set $S\subset \mathbb R^d$,
as introduced in Example \ref{ex:BR}. The extremal coefficient of the spatial average then is
\begin{equation}
  \label{theta_BR} \log  \theta_0^{\text{avg}} = -\frac{\int_{S}\int_{S} A(s)A(t)\gamma(s,t) \mathrm{d}s \mathrm{d}t}{4 \left\{\int_{S}A(s)\mathrm{d}s\right\}^2}  .
\end{equation}
    Let $d=1$ and $S=[0,T]$, $T>0$, and consider the popular power variogram model, namely
  $\gamma(s,t) = | (s-t)/\lambda |^\alpha$ for some $\alpha\in(0,2], \lambda >0$.
  In this case we obtain
  $$ \theta_0^{\text{avg}} = \exp\left\{- \frac{T^\alpha}{{ 2}\lambda^\alpha(\alpha+1)(\alpha+2)}\right\}. $$
\end{example}

\subsection{Multivariate limiting distributions of aggregated data} \label{subsec:multiv-asymptotics}

In the previous section we derived the univariate tail distribution of data 
aggregated through a functional $\ell$. In applications we often observe data through several different functionals, e.g., the integrals over
not necessarily disjoint areas. The consistency of return level estimates discussed
in the introduction has even more important implications when different
risk functionals are applied to the data. The univariate tail of each aggregation 
could be estimated separately, but the dependence between the tails would not be captured.
We consider therefore arbitrary positively homogeneous, uniformly continuous 
functionals $\ell_1, \ldots, \ell_L: C(S) \to \RR$, and we aim at describing the
multivariate tail behavior of the vector $(\ell_1(X),\dots, \ell_L(X))$.

The proof of the following theorem can be found in Appendix \ref{thm2:proof}.

\begin{theorem}\label{thm2}
 Let $\ell_1,\ldots,\ell_L$ be positively homogeneous and uniformly continuous 
 functionals on $C(S)$. Further, assume that \eqref{eq:x-conv} and \eqref{eq:decomp-a}
 hold. If $\xi \leq 0$, the spectral functions $W$ belonging to the process
 $Z$ in \eqref{eq:x-conv} are assumed to be strictly positive. Then, for 
 $\xi \neq 0$ and $\xi x_1, \ldots ,\xi x_L > 0,$
 \begin{align} \label{eq:conv-ell}
    \lim_{t \to \infty} t \: \PP \left[ \exists j\in\{1,\dots, L\}:\frac{\ell_j(X)}{a(t)\ell_j\{A(s)\}} > x_j \right]
    =\EE\left(\bigvee_{j=1}^L \left[ \frac{\ell_j\{W(s)^\xi A(s)\}}{|x_j|\ell_j\{A(s)\}}\right]^{1/\xi}\right).
 \end{align}
 For $\xi=0$, we further require that \eqref{eq:decomp-a-gumbel} holds and
 that the functionals $\ell_j$, $j=1,\ldots,L$, are linear. In this case, 
 for any $x_1,\ldots,x_L \in \RR$,
 \begin{align} \label{eq:conv-ell-gumbel}
   \lim_{t \to \infty} t \: \PP \left[\exists j\in\{1,\dots, L\}:\frac{\ell_j(X) - \ell_j\{b_s(t)\}}{a(t)\ell_j\{A(s)\}} > x_j \right] 
      = \EE\left( \bigvee_{j=1}^L \exp\left[ - x_j + \frac{\ell_j\{\log [W(s)]  A(s)\}}{\ell_j\{A(s)\}}\right] \right).
 \end{align}
\end{theorem}
\medskip

The above theorem states that the vector $(\ell_1(X),\dots, \ell_L(X))$ of aggregations
is in the max-domain of attraction of the multivariate max-stable distribution
with exponent measure given by the right-hand side of \eqref{eq:conv-ell} or \eqref{eq:conv-ell-gumbel}, respectively. For the $j$th margin, for $\xi \neq 0$, the scale of the Weibull or Fr\'echet distribution is 
$\{\theta_\xi^{\ell_j}\}^\xi$, and for $\xi=0$ the location parameter of 
the Gumbel distribution is $\log \theta_0^{\ell_j}$. This recovers the univariate results in Theorem \ref{thm1}. For details on multivariate domains of attraction and exponent measures, see \citet[Chapter 5]{resnick08}.
In general this max-stable distribution will not be available in closed form, but for the purpose
of evaluating risk regions for $(\ell_1(X),\dots, \ell_L(X))$, the exponent measure can be
approximated by Monte Carlo methods. In the following important special case, we can compute
the multivariate distribution explicitly.

\begin{example}\label{ex_BR_mult}
  Consider the same framework as in Example \ref{ex1}, namely $S\subset \mathbb R^d$ compact, $\xi=0$ and $X$ is in the max-domain of attraction of Brown--Resnick process with spectral
  functions $W$. Suppose that 
  for all $j=1,\dots, L$, the functional $\ell_j$ is the spatial average over the compact region 
  $A_j\subset S$, respectively. Since $W$ is log-Gaussian in this case, 
  the random vector 
  \begin{align}\label{BR_mult}
    \big(\ell_1\{\log[W(s)] A(s)\} / \ell_1\{A(s)\}, \dots ,\ell_L\{\log [W(s)] A(s)\}/ \ell_L\{A(s)\}\big),
  \end{align}
  is multivariate Gaussian, and its variogram matrix $\Gamma \in \mathbb R^{L\times L}$ can
  be computed explicitly; see Appendix \ref{sec:HR}.
  The exponent measure in \eqref{eq:conv-ell-gumbel} therefore corresponds
  to a $L$-variate H\"usler--Reiss distribution with dependence matrix $\Gamma$
  whose $j$th margin has a Gumbel distribution with location parameter 
  $\log \theta_0^{\ell_j}$ given in \eqref{theta_BR}. 
\end{example}

\section{Statistical Inference}\label{sec:inference}

\subsection{Setting}\label{subsec:setting} 

We suppose that we observe independent data $X_1,\dots, X_n$, $n\in \mathbb N$, 
from the process $X = \{X(s): s\in S\}$, 
but only through the aggregation functionals $\ell_j$ satisfying the conditions
from Theorem \ref{thm2}. The observations are therefore $L$-dimensional and of 
the form 
$$ \left(\ell_1(X_i), \dots, \ell_L(X_i)\right), \quad i = 1,\dots, n.$$
Making use of the limit results in Theorems~\ref{thm1} and \ref{thm2}, we aim to
infer the extremal behavior of the whole process from the observed aggregated
data. This requires estimation of both the marginal tail behavior and the 
extremal dependence of $X$. Naturally, further assumptions are needed to 
render this problem well-defined. 

We suppose that the process $X$ is in the functional max-domain of attraction
of a max-stable process $Z$ as in \eqref{eq:x-conv} with marginal distributions
of $Z(s)$ of the form~\eqref{gev} for all $s\in S$. A natural and fairly 
general assumption on the marginal distributions of $X$ is to belong to a 
location-scale family, i.e., for some distribution function $F$ and 
continuous $A: S \to (0,\infty)$, $B: S \to \mathbb R$,
$$ \Pr\{X(s) \leq x\} = F\left\{ \frac{x-B(s)}{A(s)} \right\},$$
for any fixed $s \in S$. Since $X(s)$ is in the max-domain of attraction of 
$Z(s)$, the distribution of $M_n(s)$ must converge to $G_\xi$ as $n \to \infty$.
In particular, $F$ must satisfy
$ \lim_{t \to \infty} F^t\{a(t) x + b(t)\} = G_\xi(x)$ for all $x\in\mathbb R$ with $\xi x \geq 0$ and
appropriate functions $a: (0,\infty) \to (0,\infty)$ and 
$b: (0,\infty) \to \RR$. This implies that the normalizing functions $a_s$ and
$b_s$ of $X(s)$ can be chosen as
\begin{equation}\label{eq:hypAandB}
  a_s(t) = A(s)a(t), \qquad b_s(t) = B(s) + A(s)b(t), \quad t \in \RR.
\end{equation}
Moreover, if $\xi \neq 0$, without loss of generality, we may assume $B(s) \equiv 0$
by the same arguments as in the proof of Theorem \ref{thm2}.

We impose a parametric structure on the marginal scale and 
location parameters, i.e., the unknown functions $A$ and $B$, respectively, 
and the extremal dependence of $X$, which is given by the exponent measure of 
$Z$. For the marginal distributions, we assume that $A$ and $B$ 
belong to parametric families of functions $\{A_{\vartheta_A}, \ \vartheta_A \in \Theta_A\}$
and $\{B_{\vartheta_B}, \ \vartheta_B \in \Theta_B\}$ where $\Theta_A$ and
$\Theta_B$ are appropriate subsets of $\RR^{k_A}$ and $\RR^{k_B}$,
respectively. For the dependence, we suppose that the probability measure
$\mathbb{P}^{(spec)}$ induced by the spectral function $W$ of the limiting 
max-stable process $Z$ belongs to a parametric class 
$\{\mathbb{P}^{(spec)}_{\vartheta_W}, \ \vartheta_W \in \Theta_W\}$ with 
$\theta_W \subset \RR^{k_W}$. Further, the joint normalization constants 
$a(t) \in (0,\infty)$ and $b(t) \in \RR$ need to be estimated for some large 
$t$.

In the following, we present two ways to estimate the complete parameter vector
$$ \vartheta = (a(t), b(t), \vartheta_A, \vartheta_B, \vartheta_W) \in (0,\infty) \times \RR \times \Theta_A \times \Theta_B \times \Theta_W,$$
based on the marginal and multivariate tail behavior of $(\ell_1(X),\ldots,\ell_L(X))$ 
as given in Theorems~\ref{thm1} and~\ref{thm2}, respectively.

\subsection{Least squares fit based on marginal estimates} \label{subsec:ls}

Throughout the rest of this section, for the sake of simplicity we assume that 
$\xi=0$ is known. Estimation for the case that $\xi$ is unknown can be performed
analogously, see Appendix \ref{app:margins}.

As a first approach, we approximate the tail of the distribution of $\ell_j(X)$ 
separately for each $j=1,\ldots,L$. From \eqref{eq:conv-ell-gumbel}, we obtain
for exceedances over sufficiently large $x \in \RR$ and $t>0$,
\begin{equation}  \label{eq:exceedance}
 \PP\left\{ \ell_j(X)  > x\right\} \approx t^{-1} \exp\left(-\frac{x - \mu_{j,t}}{\sigma_{j,t}}\right),
\end{equation}
or, equivalently, for block maxima,
\begin{equation} \label{eq:blockmaxima}
 \left[\PP\left\{ \ell_j(X) \leq x\right\} \right]^t \approx  \exp\left\{ - \exp\left( - \frac{x - \mu_{j,t}}{\sigma_{j,t}}\right) \right\},
\end{equation}
where the location parameters $\mu_{j,t}$ and the scale parameters
$\sigma_{j,t}$, $j=1,\ldots,L$, are given by
\begin{align} \label{eq:mu}
 \mu_{j,t}(\vartheta) &= \ell_j\{A_{\vartheta_A}(s)\} \cdot \left\{ b(t) + a(t) \log \theta_0^{\ell_j}(\vartheta) \right\} + \ell_j\left\{B_{\vartheta_B}(s)\right\},\\
 \label{eq:sigma}
 \sigma_{j,t}(\vartheta) &= a(t) \cdot \ell_j\{A_{\vartheta_A}(s)\},
\end{align}
respectively. While the asymptotic behavior of $\mu_{j,t}$ and $\sigma_{j,t}$
is uniquely determined by Equation~\eqref{eq:exceedance}, additional assumptions
on $A(s)$ and $B(s)$, such as $\ell_1\{A(s)\}=1$ and $\ell_1\{B(s)\}=0$, are necessary
to ensure the identifiability of $a$, $b$, $A$, $B$ and $ \theta_0^{\ell_j}$ from
Equations \eqref{eq:mu} and \eqref{eq:sigma}.

For large $t$, estimates $\hat \mu_{j,t}$ and $\hat \sigma_{j,t}$ for $\mu_{j,t}$ 
and $\sigma_{j,t}$ can be obtained using well-known techniques from univariate 
extreme value statistics based on peaks-over-threshold or block maxima approaches,
for instance. Here, the value of $t$ is typically closely related to the choice
of the threshold and the block size, respectively; see Appendix \ref{app:margins}
for details. Based on these estimates, we define a weighted least squares estimator
$$ \hat \vartheta_{\LS} = 
     \mathrm{argmin}_\vartheta \sum_{j=1}^L v_{j} \cdot \left\{\hat \mu_{j,t} - \mu_{j,t}(\vartheta)\right\}^2 
                                                         + w_j \cdot \left\{\hat \sigma_{j,t} - \sigma_{j,t}(\vartheta)\right\}^2,$$
where $v_j, w_j \geq 0$, $j=1,\ldots,L$, are appropriate weights and 
$\mu_{j,t}(\vartheta)$ and $\sigma_{j,t}(\vartheta)$ are given by Equations 
\eqref{eq:mu} and \eqref{eq:sigma}, respectively. A possible choice for 
$1/v_j$ and $1/w_j$ are the variances of the estimators $\hat \mu_{j,t}$ and 
$\hat \sigma_{j,t}$.
\medskip

\subsection{Censored likelihood for the joint tail behavior}\label{subsec:cens}

Alternatively, we can estimate $\vartheta$ making use of the multivariate tail
behavior of the whole vector $(\ell_1(X), \ldots, \ell_L(X))$. For sufficiently
large $x_1,\ldots,x_L \in \RR$ and $t>0$, by Theorem \ref{thm2},
\begin{equation*} \label{eq:normalizedtail}
 \PP\left\{ \exists j\in\{1,\dots, L\}: \frac{\ell_j(X) - \mu_{j,t}}{\sigma_{j,t}} > x_j\right\} \approx t^{-1} V_\vartheta(x_1,\ldots,x_L),
\end{equation*}
where 
\begin{equation}\label{eq:aggexponentmeasure}  
     V_\vartheta(x_1,\ldots,x_L)
   = \Es_{\vartheta_W} \left(\max_{j=1}^L \, \exp\left[- x_j - \log\{\theta_0^{\ell_j}(\vartheta)\} + \frac{\ell_j\{\log [W(s)]  A_{\vartheta_A}(s)\}}{\ell_j\{A_{\vartheta_A}(s)\}}\right]\right),
\end{equation}
is the exponent measure of a max-stable vector with standard Gumbel margins.
Thus, the parameter vector $\vartheta$ can be estimated by a censored likelihood
approach. Define a vector $u = (u_1,\dots u_L)$ whose $j$th element $u_j\in \mathbb R$
is a suitably high marginal threshold for $\ell_j(X)$, such as its empirical
$(1-1/t)$-quantile, and let $\mathcal K_i = \{j=1,\dots L: \ell_j(X_i) > u_j\}$. Denoting 
the normalized thresholds and data by $\tilde u = (\tilde u_1, \dots, \tilde u_L)$ and 
$Y_i = (Y_{i1}, \dots , Y_{iL})$ with
$$ \tilde u_j = \frac{u_j - \mu_{j,t}}{\sigma_{j,t}}, \qquad  Y_{ij} = \frac{\ell_j(X_i) - \mu_{j,t}}{\sigma_{j,t}}, \qquad j=1,\dots, L,$$
respectively, we let $\hat \vartheta_{\cens}$ be the $\mathrm{argmax}$ of the log-likelihood
\begin{equation}\label{eq:pplikelihood} 
  (n - |\mathcal I|)\log\left\{ 1- \frac 1 t V_\vartheta(\tilde u)\right\} 
   + \sum_{i\in\mathcal I} \log \left( \left[\prod_{j\in \mathcal K_i} \frac{1}{a(t) \cdot\ell_j\{A_{\vartheta_A}(s)\}} \right] \cdot (-1) \cdot \frac 1 t V_{\vartheta, \mathcal K_i}(Y_i)\right),
\end{equation}
where $\mathcal I = \{i = 1,\ldots,n: \ \ell_j(X_i) > u_j \text{ with } j=1,\ldots,L\}$ 
and $V_{\vartheta, \mathcal K_i}$ are the partial derivatives of $V_\vartheta$ 
in directions $\mathcal K_i$. By the homogeneity of $V_\vartheta$, it can be seen 
that the likelihood~\eqref{eq:pplikelihood} asymptotically does not depend on the 
specific choice of $t$, but only on the $u_1,\ldots,u_L$. This likelihood corresponds
to multivariate threshold exceedances and their approximation by Pareto processes 
\citep[cf.,][]{thi2015a}. The censoring of the exponent measure $V_\vartheta$ reduces 
possible bias for observations below the marginal threshold that might not yet have
converged to the limit model; see \cite{wadsworth-tawn14}.

\section{Simulation of Extreme Events}\label{sec:simu}

Environmental risk assessment is often based on rare event simulation of scenarios
with long return periods. Two kinds of simulations are typically required: 
unconditional simulations of a given or fitted model capturing the spatial extent 
and the variability of possible extreme events; and simulations at points of 
interest conditional on a particular event that was only observed at different
locations or scales. Conditional and unconditional simulations have for instance
been studied for max-stable processes \citep{DEMR13, dom2016a} and for threshold 
exceedances \citep{thi2015a, def2017}.

In this section, we discuss how the multivariate result in Theorem \ref{thm2} allows
us to perform these two kinds of simulations for extreme events of the process $X$. 
We assume that the process $X$ satisfies the assumptions of Theorem \ref{thm2} 
for known normalizing functions $a_s$ and $b_s$ with representation \eqref{eq:hypAandB}, extreme value index 
$\xi \in \RR$, and known distribution of the spectral process $W$. For simplicity, 
we again restrict to the case $\xi=0$, but the procedure can be adapted for the case
$\xi\neq 0$.

In order to simulate $X$ at a finite number of locations $s_1,\ldots,s_K \in S$,
we artificially augment the vector of functionals to $(\ell_1(X), \ldots, \ell_L(X),
\ell_{L+1}(X), \dots, \ell_{L+K}(X))$, where $\ell_{L+k}(X) = X(s_k)$ is the point 
evaluation at location $s_k$, $k=1,\ldots,K$. We apply Theorem \ref{thm2} to this 
augmented vector to obtain
$$ \lim_{t \to \infty} t \: \PP \left[ \exists j\in\{1,\dots, L+K\}: \frac{\ell_j(X) - \mu_{j,t}}{\sigma_{j,t}} > x_j \right] = \EE\left\{ \bigvee\nolimits_{j=1}^{L+K} \exp(-x_j + \log\Psi_j) \right\}, $$
where
$$ \Psi_j =  \exp\left[\frac{\ell_j\{\log [W(s)] A(s)\}}{\ell_j\{A(s)\}}   - \log \theta_0^{\ell_j} \right], \quad j=1,\ldots,L+K, $$
and $\mu_{j,t}$ and $\sigma_{j,t}$, $j=1,\ldots,L+K$, $t > 0$, are defined
as in Equations \eqref{eq:mu} and \eqref{eq:sigma}, respectively. In other words,
$(\ell_1(X),\ldots, \ell_{L+K}(X))$ is in the-max-domain of attraction 
of a max-stable distribution with standard Gumbel margins and spectral vector
$(\Psi_1,\ldots,\Psi_{L+K})$. 

In the framework of conditional simulation of an extreme event, the aggregated
data $\ell_1(X)=y_1,\ldots,\ell_L(X)=y_L$, are observed and one of them, say 
$\ell_{J}(X)$, is assumed to be large. Reformulating Theorem \ref{thm2} in terms
of threshold exceedances, we obtain the convergence in distribution
\begin{equation} \label{eq:pareto}
 \mathcal{L}\left[ \left\{\frac{ \ell_j(X) - \mu_{j,t}}{\sigma_{j,t}}\right\}_{j=1}^{L+K} \, \bigg| \, \frac{\ell_{J}(X) - \mu_{J,t}}{\sigma_{J,t}} > 0\right] 
 \longrightarrow \mathcal{L}(U + \log \Psi^{(J)}),
\end{equation}
as $t \to \infty$, where $U$ is a standard exponential random variable and, 
independently of $U$, $\Psi^{(J)}$ is a $(L+K)$-dimensional random vector
with the transformed distribution $P_J$ given in \citet[Proposition 1]{dom2016a}.
For most popular models in spatial extremes, $P_J$ can be simulated easily. 
Using approximation \eqref{eq:pareto} with $u = (y_J - \mu_{J,t})/\sigma_{J,t} > 0$
for some large $t$, we can perform conditional simulation of the vector 
$(X(s_1), \dots, X(s_K))$ in the following way.
\begin{enumerate}
 \item[(i)] Simulate a realization $(\psi_{L+1},\ldots,\psi_{L+K})$ 
       of the conditional distribution of $(\Psi^{(J)}_{L+1},\ldots,\Psi^{(J)}_{L+K})$ 
       given that $\log \Psi^{(J)}_j = (y_j -\mu_{j,t})/\sigma_{j,t} - u$, $j=1,\ldots,L$.
 \item[(ii)]  As a conditional realization of $(X(s_1),\ldots,X(s_K))$, return $x=(x(s_1),\ldots,x(s_K))$ with
       \begin{align}\label{eq:x}
       x(s_k) = a_s(t) \cdot (u + \log \psi_{L+k}) + b_s(t), \qquad k=1,\ldots,K.
       \end{align}           
\end{enumerate}
For unconditional simulation, one is typically interested in extreme events 
in the sense that at least one of the functionals $\ell_1(X), \dots, \ell_L(X)$ 
exceeds a high threshold. Therefore, we replace the conditioning event in \eqref{eq:pareto} 
by $\max_{j=1,\dots,L} \{\ell_{j}(X) - \mu_{j,t}\}/\sigma_{j,t} > 0$, such that the
vector $\Psi^{(J)}$ in the limiting law in \eqref{eq:pareto} becomes a vector 
$\Psi^{(\max)}$ that is normalized with respect to the maximum of its first 
$L$ components \citep[cf.,][]{dom2016}. Noting that $\Psi^{(\max)}$ can be 
generated by rejection sampling \citep[cf.,][]{def2017}, we can adapt the 
conditional simulation procedure to obtain an unconditional extreme sample. 
Indeed, it suffices to let $(\psi_{1},\ldots,\psi_{L+K})$ in (i) be a realization 
of the unconditional distribution of  $\Psi^{(\max)}$, and to replace the constant
$u$ in Equation \eqref{eq:x} in (ii) by a realization of the standard exponential 
distribution $U$.

In order to perform conditional and unconditional simulation, the multivariate 
tail behavior of the vector $\ell(X)$ in the sense of Theorem \ref{thm2} is 
required. For our running example of a limiting Brown--Resnick process, the 
following makes this explicit.

\begin{example}\label{ex_joint}
As in Example \ref{ex_BR_mult}, let $S\subset \mathbb R^d$ be compact, $\xi=0$ 
and $X$ is in the max-domain of attraction of a Brown--Resnick process. The 
aggregation functionals $\ell_j$ are spatial averages over compact regions 
$S_j\subset S$, $j = 1,\dots, L$, or point evaluations $\ell_{L+k}(X) = X(s_k)$ 
at locations $s_k \in S$, $k=1,\dots, K$. 
The vector $(\ell_{1} (X), \dots, \ell_{L + K} (X))$ then satisfies 
the assumptions of Theorem \ref{thm2}, and it is in the max-domain of attraction 
of a multivariate H\"usler--Reiss distribution with dependence matrix
\begin{align*}    
  \Gamma = \left(\begin{array}{cc}\{\Gamma_{jk}\}_{j,k} & \{\Gamma_{jq}\}_{j,q} \\ \{\Gamma_{pk}\}_{p,k} & \{\Gamma_{pq}\}_{p,q}\end{array}\right), \quad 
  \left\{ \begin{array}{l}
    j,k = 1,\dots, L, \\
    p,q = L+1,\dots, L+K.
  \end{array}\right.
\end{align*}
The entries of the four sub-matrices and the explicit form of the exponent 
measure are given in Appendix \ref{sec:HR}. In
this case, the above algorithms essentially reduce to conditional and 
unconditional simulation of Gaussian processes.
\end{example}

\section{Application: downscaling extremes}\label{sec:application}

\subsection{Statistical downscaling} \label{sec:downscaling}

Environmental data can be classified into two broad categories. On the one hand,
station measurements are obtained through direct observation of the physical quantity. 
This type of data refers to a precise location in space, but it may suffer from
inhomogeneities between stations due to varying record lengths and differences 
between measurement instruments, and, moreover, it usually has a sparse spatial 
coverage. Gridded databases, for instance generated by climate models, on the other hand, cover a
large region or even the entire globe, but at a coarse scale where data points 
can be considered as
an aggregation of the physical variable. 

Understanding the link
from these gridded data to point measurements is an important area of research in
environmental sciences called downscaling. Besides dynamical downscaling procedures
based on the solution of partial differential equations describing the physical 
processes, a large number of downscaling techniques relying on the statistical
relationship between variables at different scales have been applied.
Most of these techniques focus on central characteristics of the distribution such
as mean and variance. In geostatistics, for instance, the so-called change of support
has been extensively studied for Gaussian processes 
\citep[cf.,][and references therein]{chilesdelfiner2012}. There are only few examples
of statistical downscaling procedures for extremes. \cite{man2010} and \cite{kal2011} follow
an approach related to univariate extreme value theory, and
\cite{bechleretal15} and \citet{oestingetal17} propose conditional simulation 
from a spatial max-stable process that has been estimated from station measurements.

Here, using the theoretical results in Section \ref{sec:theory}, we extend the idea 
of changing the support of a stochastic process $X$ to the context of extremes, basing 
inference only on aggregated observations $\ell_{1}(X), \ldots, \ell_{L}(X)$. These might
come from gridded data sets, as in our case, supposing that the grid values represent an
aggregation of the underlying physical quantity. If additional station measurements 
$X(s_{1}), \dots, X(s_{K})$ are available, they can also be used. Outputs of the
method will be return level estimates at point locations, as well as unconditional and 
conditional simulations of rare events in the region $S$. The method allows for the
estimation of marginal characteristics such as return levels at point locations, as well as 
unconditional and conditional simulations of rare events on the entire region $S$.
 
\subsection{Application to extreme temperature in the South of France}\label{sec:downscalingsouthfrance}

\begin{figure} 
\begin{center}
\begin{tabular}{ccc}
\hspace*{-3em} {\includegraphics[width=0.4\textwidth]{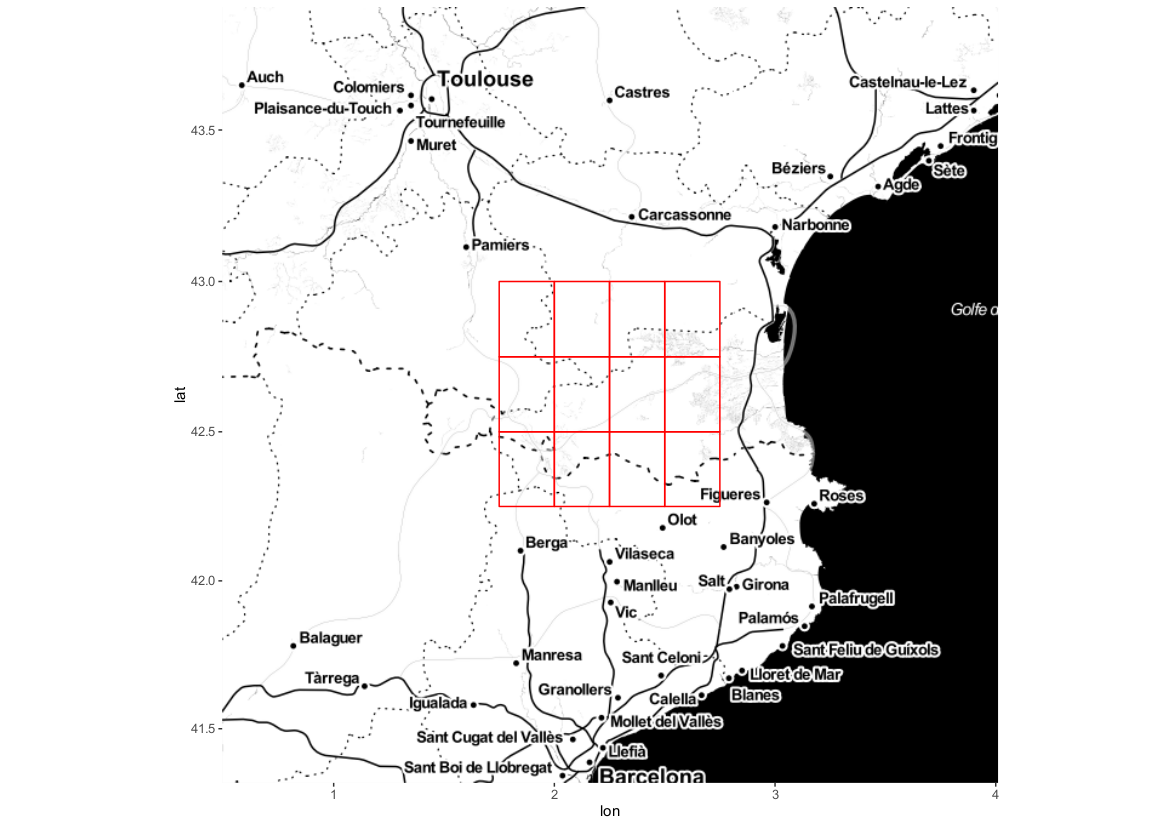}} &
\hspace*{-3em}  {\includegraphics[width=0.4\textwidth]{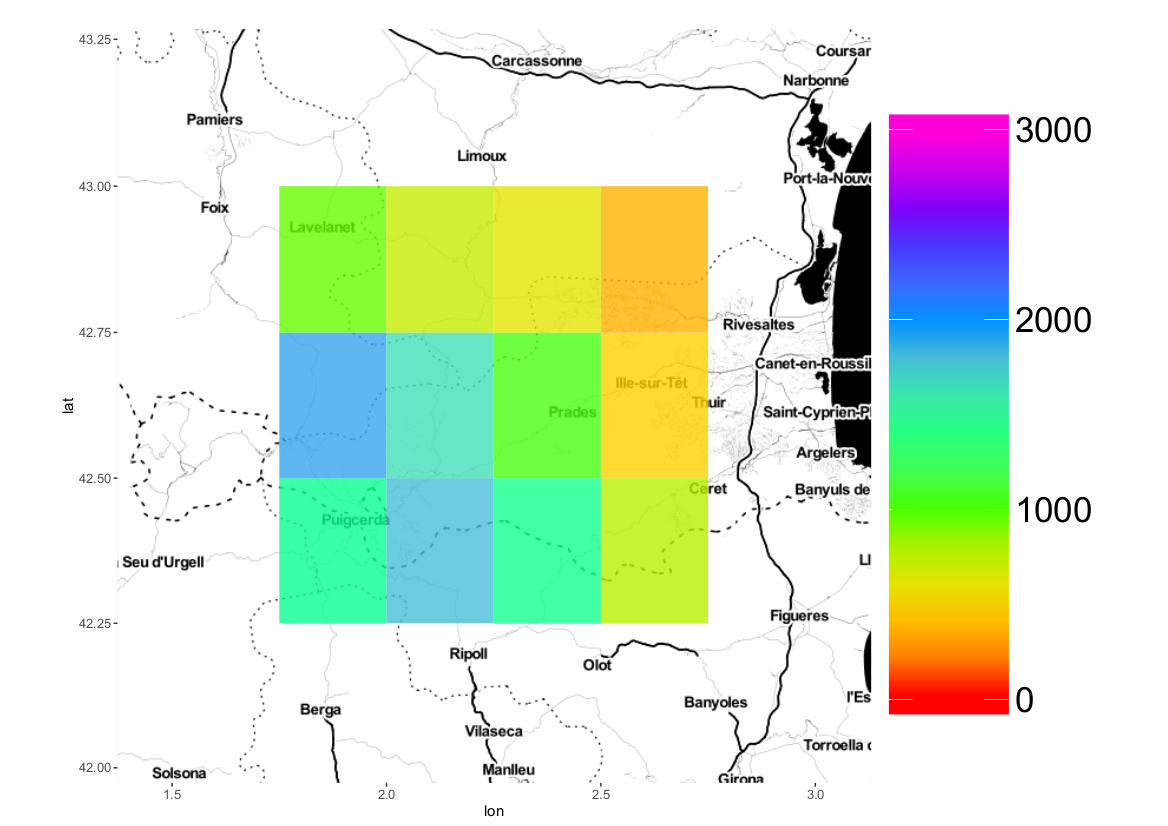}} & \hspace*{-2em}  {\includegraphics[width=0.39\textwidth]{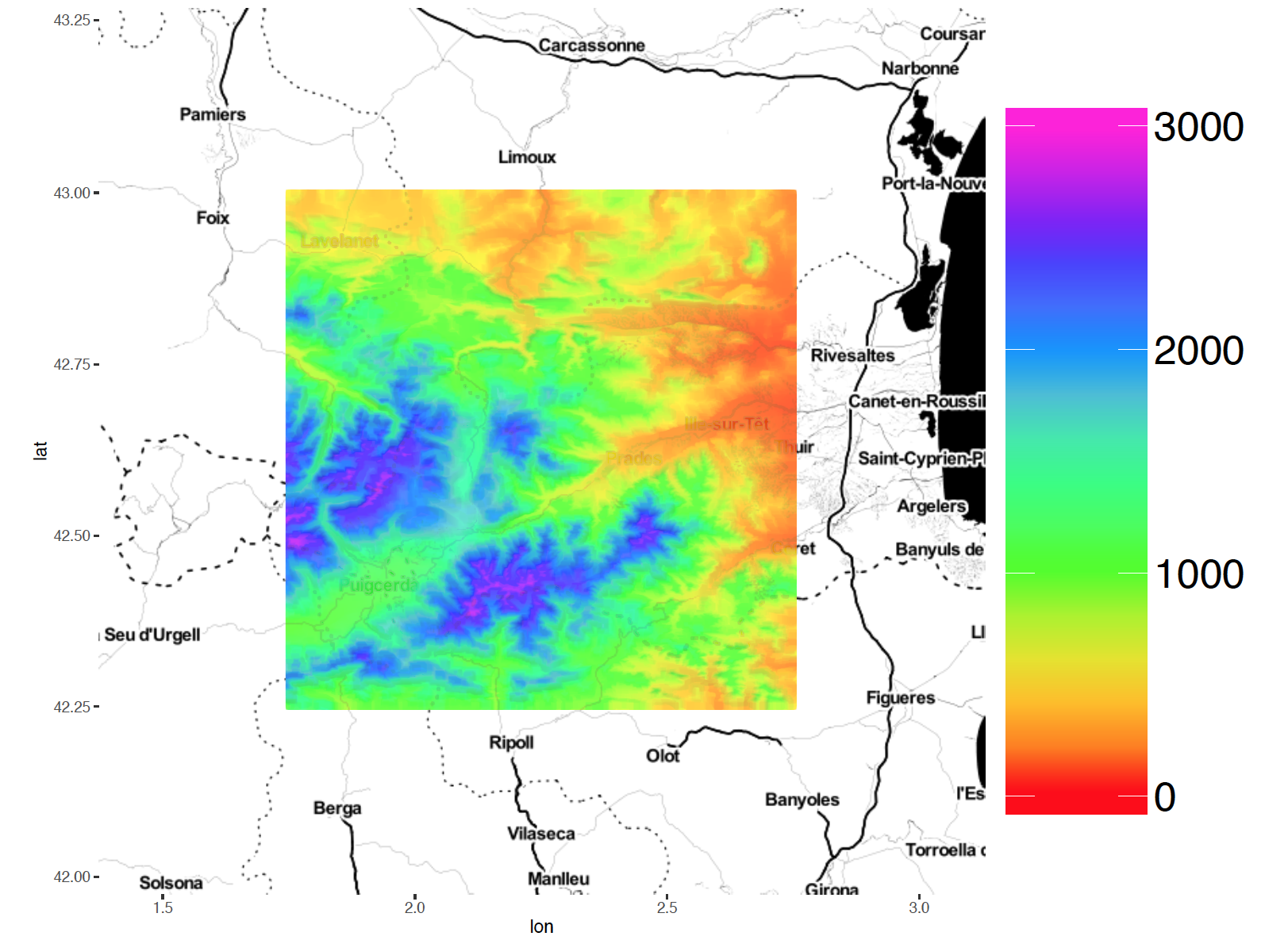}} 
\end{tabular}
\end{center} 
 \caption{The study region consisting of 12 grid cells in the south of France (left),
          mean altitude within each cell (middle) and elevation in the region.} \label{fig:region}
\end{figure}

We apply our downscaling procedure to daily temperature maxima in Europe from the e-obs data set
\citep{Haylock2008}, which covers the period from $1950$ to $2016$ with a $0.25^{\circ}$ grid 
resolution. To avoid potential temporal non-stationarity, we restrict the study to the summer 
period, i.e., July and August. Our study region $S$ is a $80\text{km} \times 80\text{km}$ 
subset of the gridded product located in the south of France, in the west of Perpignan; see
Figure~\ref{fig:region}. The region is mountainous and thus altitude appears to be a
natural covariate for our model. The underlying spatial process of temperatures is denoted by
$\{X(s): s\in S\}$, and the observations $(\ell_1(X_i), \dots, \ell_L(X_i))$ on day $i$ can be considered as the spatial averages 
over the $L=12$ cells in $S$, where $i = 1,\dots, n$, and $n$ is the number of days in the given
time span of $67$ years. The null hypothesis 
that the marginal tails of the aggregated data are in the Gumbel domain of attraction cannot be
rejected, and we thus assume in sequel that $\xi=0$. 

Throughout we assume the same setting as in Section \ref{subsec:setting}, namely that the marginal
distributions of $X(s)$ belong to a location-scale family for all $s\in S$ parameterized through the
functions
$$
\begin{array}{lll}
A(s) & =  1, \qquad B_{\vartheta_B}(s) & =  b_0 + b_1 \times \mathrm{alt}(s) + b_2 \times \mathrm{lon}(s) + b_3 \times \mathrm{lat}(s),
\end{array}
$$
where $\mathrm{alt}(s)$, $\mathrm{lon}(s)$ and $\mathrm{lat}(s)$ denote the altitude, longitude and latitude at location $s \in S$, respectively. We further suppose that $X$ is in the functional 
max-domain of attraction of a max-stable process $Z$ belonging to a parametric family $\{Z_{\vartheta_W}: 
\, \vartheta_W \in \Theta_W\}$, for which we consider the Brown--Resnick processes introduced 
in Example \ref{ex:BR}, parameterized by $ \vartheta_W = (\alpha,\lambda,\eta,a)$ for the anisotropic power variogram
\begin{equation*}
\gamma(s_1,s_2) = \left\| \frac{\Omega (s_1 - s_2)}{\lambda} \right\|^\alpha,  \quad s_1,s_2 \in S,
\end{equation*}
with $0 < \alpha \leqslant 2,\lambda>0$ and anisotropy matrix
\begin{equation*}
\Omega = 
\left[\begin{array}{cc}\cos \eta & -\sin \eta \\ a \sin \eta & a \cos \eta\end{array} \right], \quad \eta \in \left(-\frac{\pi}{2}; \frac{\pi}{2}\right], \quad a > 1.
  \end{equation*}

In Sections \ref{subsec:ls} and \ref{subsec:cens} we discussed two approaches to
estimate the parameters of this model, namely least squares estimation based on
univariate location and scale estimates, and censored likelihood estimation for
multivariate threshold exceedances. The formulas required for the implementation of these
approaches have been derived in Sections \ref{sec:theory} and \ref{sec:inference} 
and in Appendix \ref{sec:HR}. For least squares estimation, 
this includes the explicit expression \eqref{theta_BR} for the univariate $\ell$-extremal 
coefficient. For censored likelihood estimation of the model parameters in \eqref{eq:pplikelihood},
we require the partial derivatives $V_{\mathcal K}$ of the exponent measure $V$, which can be
obtained as in \citet[Section 4.3.2]{asa2015}; see Appendix \ref{sec:HR} for more details. 
In order to assess the effectiveness and to compare the efficiency of the
two methods, in Appendix \ref{sec:simustudy} we perform a 
simulation study with a setup similar to this application. It turns out that the censored
likelihood approach is significantly more efficient since it uses the full information on 
extremal dependence.

The parameters of our model for temperature extremes are therefore fitted using the censored
likelihood procedure based on all observations where at least one component exceeds its respective 
 empirical $0.98$ quantile. To avoid possible temporal dependence we keep only 
observations that are at least $5$ days apart, yielding a set of $114$ events. The parameter
estimates are displayed in Table \ref{tab:param} where standard deviations are obtained using
a jackknife procedure with $19$ blocks of size $6$; censored maximum likelihood is performed
repeatedly with one block left out.

\begin{table}
\begin{center}
\begin{tabular}{c c c c c c c c c c}
& $a_n$ & $b_n$ & $b_1$ & $b_2$ & $b_3$ & $\alpha$ & $\lambda$ & $\eta$ & $a$\\ \hline
Estimate &$1.90$  & $35.53$  & $4.51$ & $-0.53$ &  $-0.20$ & $0.90$ & $6.42$ & $-0.08$ & $1.14$\\ 
Standard deviation &  $0.06$ & $0.27$ & $0.14$ & $0.26$ & $0.28$ & $0.07$ & $0.51$ & $0.22$ & $0.08$
\end{tabular}
\end{center}
\caption{Estimated parameters and standard deviations for the temperature downscaling model. 
         Standard deviations are computed using a block jackknife with $19$ blocks of size $6$.}  \label{tab:param}
\end{table}

We assess the model fit in the diagnostic plots shown in Appendix 
\ref{app:assessment}. First, we check the marginal distributions implied by the fitted linear model by comparing 
them in quantile-quantile plots to the observations; see Figure~\ref{fig:qqplot}. The 
model provides a good fit for most stations and the quantiles of the fitted model generally remain
in the confidence bounds obtained by parametric bootstrap. For a small number of stations, 
the model slightly over-estimates return levels.

Verification of the dependence structure is based on a graphical comparison of the pairwise extremogram \citep{davis2009}
from the fitted multivariate H\"usler--Reiss model to its empirical counterpart based on the gridded observations. 
The extremogram values were significantly larger than zero for increasing thresholds and stable around the
empirical $0.98$ quantile, validating the asymptotic
dependence model. Figure~\ref{fig:dep} shows that the fitted 
variogram model successfully captures the major trend of the cloud of points. The effect of spatial anisotropy 
seems to be rather weak, which is also reflected in the parameter estimate for~$a$ close to $1$.

\begin{figure}
\begin{center}
\begin{tabular}{ccc}
$50$ years & $100$ years \\
\includegraphics[width=0.37\textwidth]{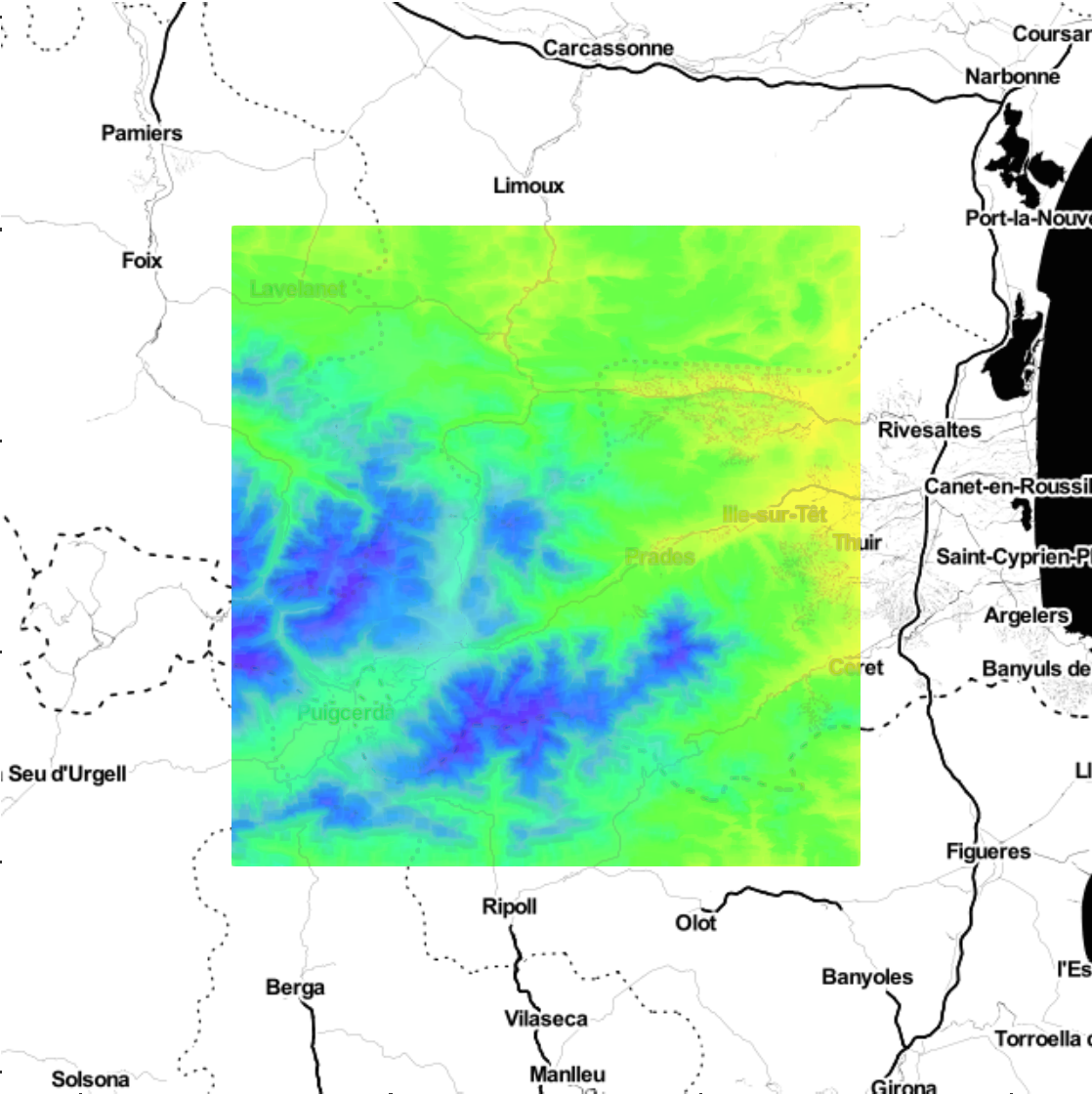}  & \includegraphics[width=0.37\textwidth]{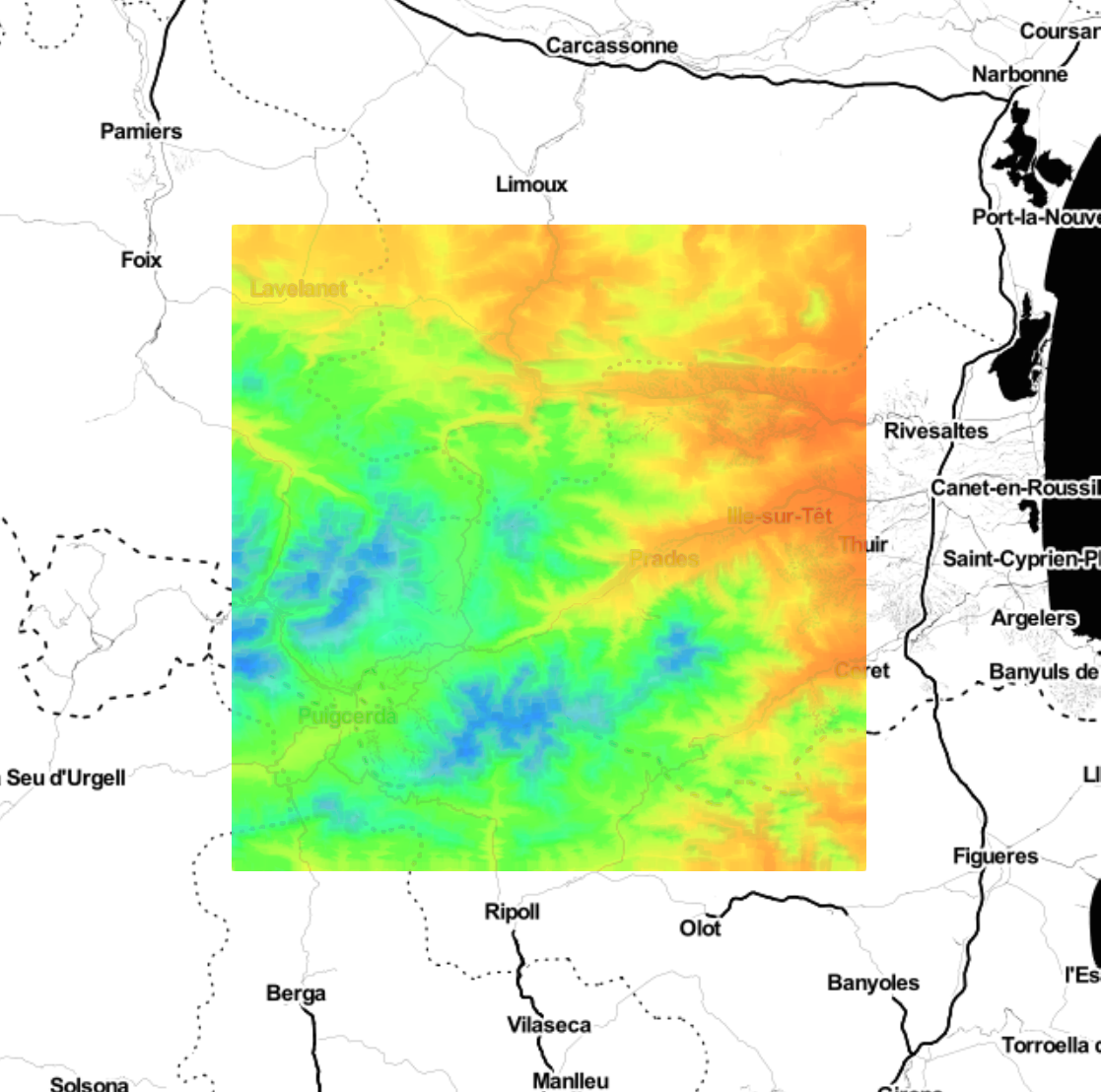} &  \includegraphics[width=0.1\textwidth]{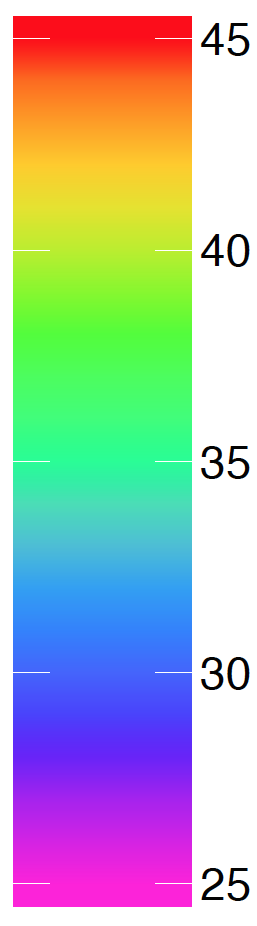} \\
\end{tabular}
\end{center}
\caption{Downscaled return levels of daily temperature maxima ($^{\circ}$C) for the $50$ (left) and $100$ (right) years return periods in the south of France at a $25  \times 25$m resolution.} \label{fig:return}
\end{figure}
 
\begin{figure}
\begin{center}
\begin{tabular}{cccc}
\hspace*{-4.5em} \includegraphics[width=0.4\textwidth]{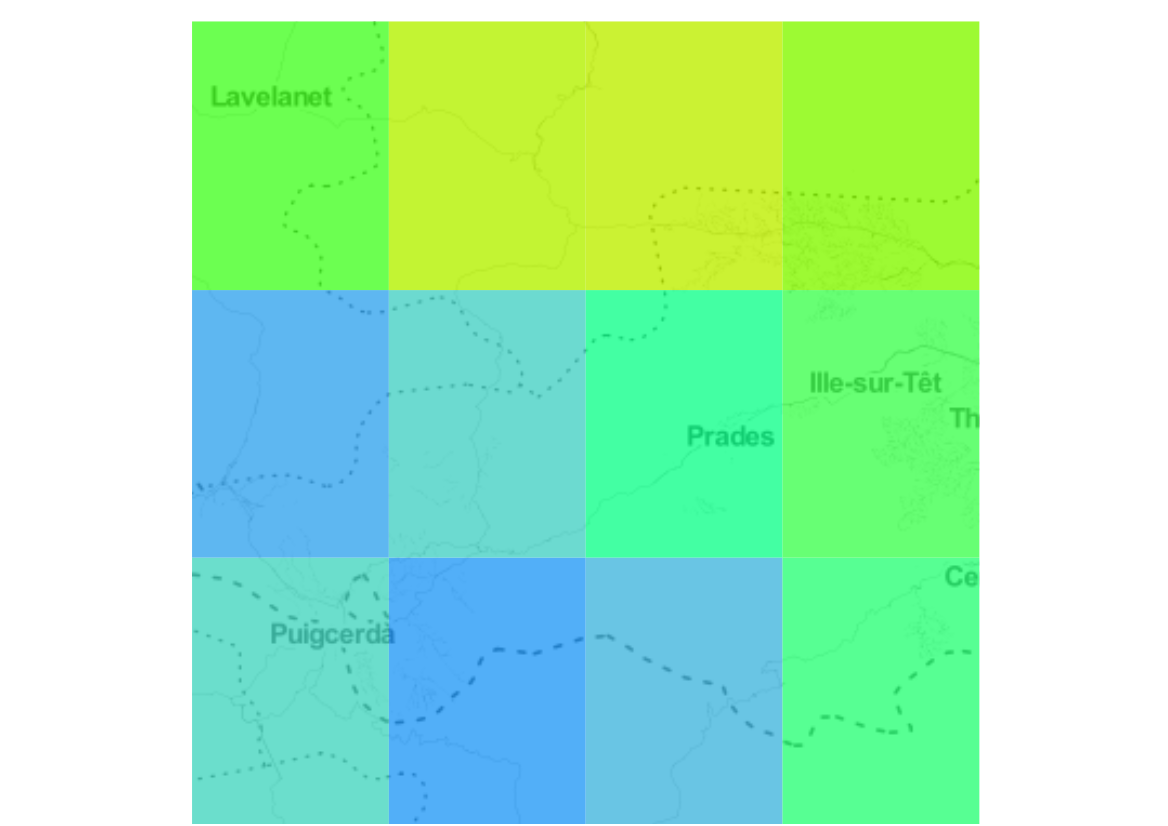} 
& \hspace*{-3.3em}\includegraphics[width=0.4\textwidth]{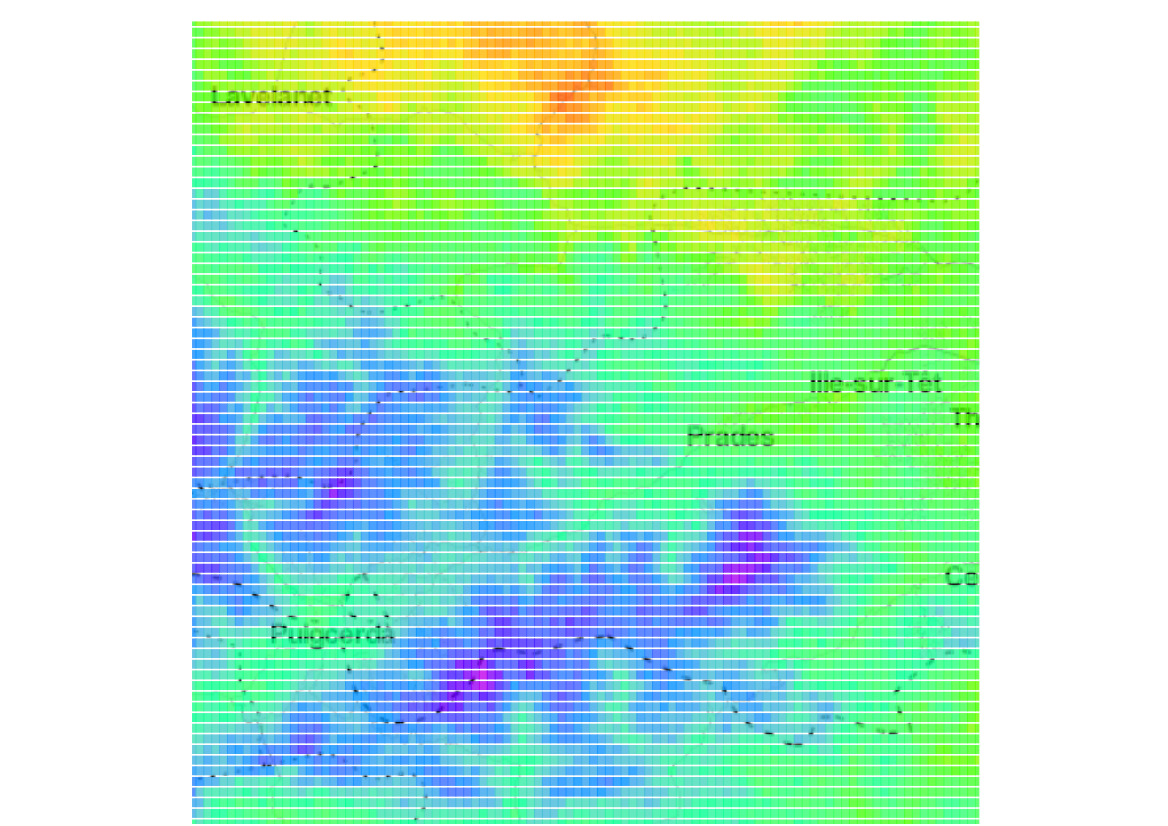}
& \hspace*{-3.3em} \includegraphics[width=0.4\textwidth]{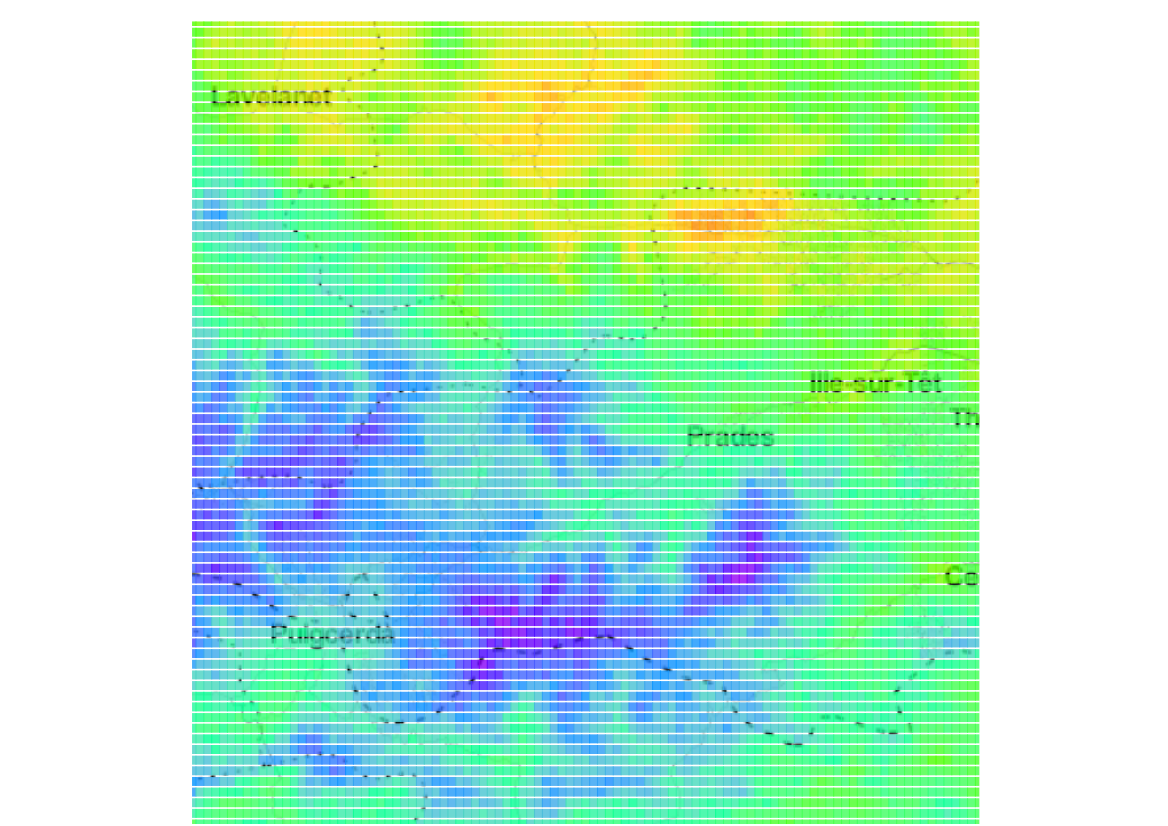}
&  \hspace*{-3em} \includegraphics[width=0.08\textwidth]{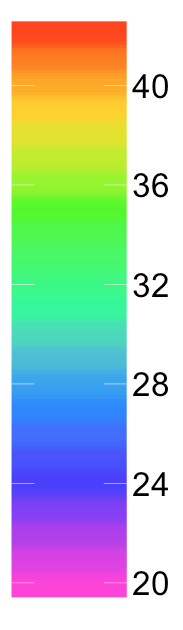} \\
\end{tabular}
\end{center}
\caption{Maximal temperature ($^{\circ}$C) on the warmest day during the $2003$ heatwave. Gridded data from the e-obs database
         \citep{Haylock2008} (left); conditional simulations with a $1 \times 1$ km resolution (center and right).} \label{fig:condsim}
\end{figure}

The fitted marginal model allows us to obtain return level maps for point locations at arbitrarily fine resolutions. 
In Figure~\ref{fig:return}, we produced such maps for the $50$ and $100$ year return periods. The full 
fitted model of marginal distributions and dependence structure further enables us to conditionally and unconditionally
generate spatial extreme events of temperature fields at both a coarse and a fine resolution grid via the simulation 
procedures described in Section \ref{sec:simu}. Figure~\ref{fig:condsim}, for instance, displays two high
resolution simulations of the temperature field conditionally on the observed aggregated temperatures during the warmest 
day of the $2003$ heatwave. The simulations show that extreme temperatures at fine resolutions 
can be remarkably larger than at a coarse scale.
Moreover, both simulations are constrained to have the same observed averages on the grid boxes, but they may
exhibit different spatial patterns. This illustrates the variability of such a heatwave and provides practitioners with a set of possible scenarios that can be used for risk assessment.

\section*{Acknowledgement}                                                   

We would like to thank Anthony C.~Davison for helpful comments and discussions, and the Swiss National Science Foundation for financial support.

\bibliographystyle{abbrvnat}
\bibliography{lit}

\appendix

\setcounter{figure}{0}
\renewcommand{\thefigure}{A\arabic{figure}}
\renewcommand{\thetable}{A\arabic{table}}

\section{Proof of Theorem \ref{thm2}}\label{thm2:proof}

\begin{proof}
Condition \eqref{eq:x-conv} implies that the exponent measure $\nu$ of $Z$, defined by
\begin{equation} \label{eq:exp-measure}
\nu(E) = \EE \left( \int_0^\infty u^{-2} {1}{\{u W(s) \in E\}} \sd u\right), \quad E \in \C_+(S),
\end{equation}
where $C_+(S)$ and $\C_+(S)$ denote the analogues to $C(S)$ and $\C(S)$ for non negative
functions, verifies
\begin{equation} \label{eq:conv-exp-measure}
 \nu(E) = \begin{cases}
           \lim_{n \to \infty} n \: \PP\left( \left\{\left[\frac{X(s) - b_s(n)}{a_s(n)}\right\}^{1/\xi}_+, \, s \in S\right] \in E\right), & \xi \neq 0,\\
           \lim_{n \to \infty} n \: \PP\left( \left[\exp\left\{\frac{X(s) - b_s(n)}{a_s(n)}\right\}, \, s \in S\right] \in E\right), & \xi=0.
          \end{cases}
\end{equation}
Closely related to the spectral process $W$, the measure $\nu$ incorporates the extremal dependence structure of $X$. 

 For the Fr\'echet case, we note that, by Proposition 1.11 in \citet{resnick08},
 $b_s(n) \equiv 0$ is a valid choice for the norming constant. Thus, Proposition 0.2 therein
 implies that the fraction $b_s(n)/a_s(n)$ converges to zero for every $s \in S$ 
 as $n \to \infty$. By the results in Subsection 9.2 in \citet{DHF06}, the
 convergence is uniform on $S$. Further, the continuous function $A$ is strictly
 positive and thus bounded away from zero on the compact set $S$. Hence, by
 \eqref{eq:decomp-a}, for any $\varepsilon > 0$, we have 
 $|a(n)^{-1} a_s(n) - A(s)| < \varepsilon A(s)$ and $|b_s(n)/a_s(n)| < \varepsilon$ 
 for all $s \in S$ and sufficiently large $n$. We thus obtain the uniform
 bound
  \begin{align*}
   \frac{X(s)}{a(n)} ={}& \frac{a_s(n)}{a(n)} \frac{X(s) - b_s(n)}{a_s(n)}
                               + \frac{a_s(n)}{a(n)} \left(\frac{b_s(n)}{a_s(n)}\right)\\
                            \geq{}& (1-\varepsilon) A(s) \left[\frac{X(s) - b_s(n)}{a_s(n)}\right]_+ - 2 (1+\varepsilon) A(s) \varepsilon,
 \end{align*}
 and analogously
 \begin{align*}
   \frac{X(s)}{a(n)} \leq{}& (1+\varepsilon) A(s) \left[\frac{X(s) - b_s(n)}{a_s(n)}\right]_+ + 2 (1+\varepsilon) A(s) \varepsilon.
 \end{align*}
 Hence, for any realization of $X$ and sufficiently large $n$, there
 exists $\Delta(X,n) \in [1-\varepsilon,1+\varepsilon]$ such that 
 $$ \sup_{s \in S} \left| \frac{X(s)}{a(n)} - \Delta(X,n) \cdot A(s) \cdot \left[\frac{X(s) - b_s(n)}{a_s(n)}\right]_+\right| \leq 2 (1+\varepsilon) \varepsilon \cdot \sup_{s \in S}  A(s).$$
 As each $\ell_j$, $j=1,\ldots,L$, is uniformly continuous, there exists a 
 function $h: (0,\infty) \to (0,\infty)$, $\lim_{\varepsilon \searrow 0} h(\varepsilon) = 0$,
 such that $\sup_{j=1,\ldots,L} |\ell_j(f) - \ell_j(g)| \leq h(\varepsilon)$ for 
 all $f,g \in C_+(S)$ such that $\|f-g\|_\infty \leq 2 (1+\varepsilon) \varepsilon \|A\|_\infty$.
 The homogeneity of each $\ell_j$ then entails
 \begin{align*}
  \frac{\ell_j(X)}{a(n)} \geq{}& \Delta(X,n) \cdot \ell_j\left\{\left[\frac{X(s) - b_s(n)}{a_s(n)}\right]_+ A(s)\right\} - h(\varepsilon)\\
                            \geq{}& (1-\varepsilon) \cdot \ell_j\left\{\left[\frac{X(s) - b_s(n)}{a_s(n)}\right]_+ A(s)\right\} - h(\varepsilon), \quad j=1,\ldots,L,
 \end{align*}
 and
 \begin{equation*}
  \frac{\ell_j(X)}{a(n)} \leq (1+\varepsilon) \cdot \ell_j\left\{\left[\frac{X(s) - b_s(n)}{a_s(n)}\right]_+ A(s)\right\} + h(\varepsilon), \quad j=1,\ldots,L.
 \end{equation*}
 With $\varepsilon \searrow 0$, for $x_1,\ldots, x_L>0$, we obtain
 \begin{align*}
    & \lim_{n \to \infty} n \: \PP\left[ \exists \, j\in\{1,\dots, L\}: \frac{\ell_j(X)}{a(n)} > x_j\right]\\
 ={}& \lim_{n \to \infty} n \: \PP\left[ \exists \, j\in\{1,\dots, L\}: \ell_j\left\{ \left[\frac{X(s) - b_s(n)}{a_s(n)}\right]_+ A(s)\right\} > x_j\right]\\
 ={}& \lim_{n \to \infty} n \: \PP\left( \exists \, j\in\{1,\dots, L\}: \ell_j\left[\left\{\left[\frac{X(s) - b_s(n)}{a_s(n)}\right]_+^{1/\xi}\right\}^\xi A(s)\right] > x_j 
                                     \right)\\
 ={}& \nu\left(\left\{f \in C_+(S): \,  \exists \, j\in\{1,\dots, L\}\text{ with }  \ell_j\left\{ f(s)^\xi A(s)\right\} > x_j \right\}\right) \\
 ={}& \EE\left(\int_0^\infty u^{-2} {1}{\{ \exists \, j\in\{1,\dots, L\}: \ell_j[\{uW(s)\}^\xi A(s)] > x_j \}} \sd u\right)
 {}={} \EE\left(\bigvee_{j=1}^L \left[ \frac{\ell_j\{W(s)^\xi A(s)\}}{x_j}\right]^{1/\xi}\right),
 \end{align*}
 where we used \eqref{eq:conv-exp-measure} and \eqref{eq:exp-measure}.
 The proof for the Weibull case is analogous.
In the Gumbel case, the integral in the proof of Theorem 2.1 in \citet{FDHZ12}
can just be replaced by the linear functionals $\ell_1,\ldots,\ell_L$ to obtain that
 \begin{align*}
    & \lim_{n \to \infty} n \: \PP\left[ \exists \, j\in\{1,\dots, L\}: \frac{\ell_j(X) - \ell_j\{b_s(n)\}}{a(n)} > x_j  \right] \\
 ={}& \lim_{n \to \infty} n \: \PP\left[ \exists \, j\in\{1,\dots, L\}: \ell_j\left\{\frac{X(s)-b_s(n)}{a_s(n)} A(s)\right\} > x_j  \right] \displaybreak[0] \\
  ={}& \lim_{n \to \infty} n \: \PP\left\{ \exists \, j\in\{1,\dots, L\}: \ell_j\left( \log\left[\exp\left\{\frac{X(s)-b_s(n)}{a_s(n)}\right\} \right] A(s)\right) > x_j \right\}  \\
 ={}& \nu\left\{f \in C_+(S): \exists \, j\in\{1,\dots, L\} \text{ with } \ell_j[\log\{f(s)\} A(s)] > x_j  \right\},
 \end{align*}
 for $x_1,\ldots,x_L \in \RR$. Using its definition in \eqref{eq:exp-measure}, 
 the exponent measure can be calculated yielding
 \begin{align*}
     & \nu\left\{f \in C_+(S):\exists \, j\in\{1,\dots, L\} \text{ with } \ell_j(\log(f(s)) A(s)) > x \right\}\\
  ={}& \EE \left( \int_0^\infty u^{-2} {1}{\left\{\exists \, j\in\{1,\dots, L\}: \log u > \frac{x_j - \ell_j[\log\{W(s)\} A(s)]}{\ell_j\{A(s)\}}  \right\}} \sd u\right)\\
  ={}& \EE\left\{ \bigvee_{j=1}^L \exp\left(\frac{x_j - \ell_j[\log\{W(s)\}A(s)]}{\ell_j\{A(s)\}}\right) \right\}.
 \end{align*}
 Replacing $x_j$ by $x_j \cdot \ell_j\{A(s)\}$ closes the proof.
\end{proof}

\section{Background and formulas related to H\"usler--Reiss distributions}\label{sec:HR}

\subsection{H\"usler--Reiss distributions}

The class of Brown--Resnick processes takes a similar role in spatial extreme value 
statistics as Gaussian processes in classical geostatistics. In order to specify 
their finite dimensional distributions, we recall a popular model in multivariate 
extreme value theory, namely the H\"usler--Reiss distribution \citep{hue1989}. An 
$m$-dimensional max-stable random vector $(Z_1,\dots, Z_m)$ with distribution function
$F_Z(x_1,\dots,x_m) = \exp\{ -V(x_1,\dots,x_m) \}$ is H\"usler--Reiss distributed 
with Gumbel margins and strictly conditionally negative definite parameter matrix 
$\Gamma \in [0,\infty)^{m \times m}$ if its exponent measure has the form
\begin{equation}\label{eq:HR}
 V(x_1,\dots,x_m) = \Es\left[ \max_{j=1,\ldots,m} \exp\left\{-x_j + Y_j - \frac12 \Var(Y_j)  \right\}\right],            
\end{equation}
for a centered Gaussian random vector $(Y_1,\dots,Y_m)$ with variogram matrix 
$\Gamma_{jk} = \Es\{(Y_j-Y_k)^2\}$, $1\leq j,k \leq m$. In this case, one 
possible choice for the covariance matrix of $Y$ is 
\begin{align}\label{HR_cov}
  \Sigma = \frac12 \left( \Gamma_{j1} + \Gamma_{k1} - \Gamma_{jk}\right)_{1\leq j,k \leq m}.
\end{align}
The exponent measure $V$ is normalized in the sense that 
$V(\infty,\dots,x_j,\dots,\infty) = \exp(-x_j)$, for any $j=1,\dots,m$. If $Z$ is a 
Brown--Resnick process associated to the variogram $\gamma$, then the distribution of
$(Z(s_1),\dots,Z(s_m))$ is H\"usler--Reiss with parameter matrix 
$\Gamma = \{\gamma(s_j,s_k)\}_{j,k=1,\dots,m}$.

For censored likelihood estimation (cf., Section \ref{subsec:cens}) of models
with H\"usler--Reiss limit, we require the partial derivatives $V_{\mathcal K}$
of $V$ in \eqref{eq:HR} with respect to any non-empty subset of variables 
$\mathcal K \subset \{1,\dots, m\}$. Let $b\in \{1,\dots,m\}$ be the number of 
components that exceed their thresholds, and, without loss of generality, let
$\mathcal K = \{1,\dots,b\}$. Based on the results in \cite{eng2014}, 
\cite{wadsworth-tawn14} and \citet[Section 4.3.2]{asa2015}, we obtain the 
representation
\begin{equation} \label{eq:brownresnickcens}
  (-1)\cdot V_{\mathcal K}(z) =  \exp (- z_1) \varphi_{b-1}(\widetilde{z}_{2:b}; \Sigma_{2:b}) \Phi_{L-b}\{\mu_{\text{C}}(z_{1:L}),\Sigma_{\text{C}}(z_{1:b})\},
\end{equation}
where $\widetilde{z} = \{(z_j - z_1) + \Gamma_{1j}/2\}_{1\leq j \leq m}$, $\Sigma$ is
as in \eqref{HR_cov} and $\varphi_{k}(\cdot, \Psi)$ and $\Phi_{k}(\cdot, \Psi)$ are 
the multivariate density and distribution function of a $k$-variate normal distribution
with covariance $\Psi$. We use the convention that $\varphi_0\equiv 1$ if $b=1$ and 
$\Phi_0\equiv 1$ if $b=m$, respectively. The mean $\mu_C$ and covariance matrix 
$\Sigma_C$ are
\begin{align*}
\mu_{\text{C}} &=  \widetilde{z}_{(b+1):m}  - \Sigma_{(b+1):m, 2:b}\Sigma_{2:b,2:b}^{-1}\widetilde{\mathrm{z}}_{2:b}, \\
\Sigma_{\text{C}} &=  \Sigma_{(b+1):m, (b+1):m} - \Sigma_{(b+1):m, 2:b}\Sigma_{2:b,2:b}^{-1} \Sigma_{ 2:b,(b+1):m}.
\end{align*}

\subsection{Explicit formulas for extremes of aggregated data}
\label{sec:exp_formula}

In the case where the underlying process $X$ is in the domain of attraction of a 
Brown--Resnick process with Gumbel margins, we can obtain explicit formulas for 
the $\ell$-extremal coefficient and the multivariate limits for certain aggregation
functionals.

Throughout this section we work with the general assumptions and notations in Section~\ref{sec:theory},
 and concentrate on the case where $\xi=0$ and the limiting process $Z$ is a Brown--Resnick
process on a compact region $S \subset \mathbb{R}^d$. We further assume that the variogram $\gamma$ as 
defined in Example~\ref{ex:BR} depends on the spatial lag $s-t$ only and we therefore write $\gamma(s-t)$ 
for $\gamma(s,t)$. Then, without loss of generality, we may assume that $G(0) = 0$ and the 
spectral function simplifies to
$$ W(s) = \exp\left\{G(s) - \gamma(s)/2\right\}, \quad s \in S.$$
We start with proving the closed form expression of the $\ell$-extremal coefficient $\theta_0^{\text{avg}}$, 
where $\ell$ is a spatial average over the region $S$; see Example~\ref{ex1}. 
Denoting $\bar{A} = \int_{S} A(s) \, \mathrm{d}s$, it follows from Theorem~\ref{thm1} that
\begin{align}\label{formula_theta_BR} 
\theta_0^{\text{avg}} & = \Es\left(\exp\left[ \frac{1}{\bar{A}} \int_{S} \left\{G(s) - \gamma(s)/2\right\}A(s)\, \mathrm{d}s \right]\right) 
= \exp \left\{\frac{\sigma_{\text{avg}}^2}{2} - \frac{1}{2\bar{A}} \int_{S} A(s)\gamma(s)\, \mathrm{d}s\right\},
\end{align}
since the integral over a Gaussian process is normally distributed with variance
$$
\sigma_{\text{avg}}^2  =  \Var \left\{\frac{\int_{S} A(s)G(s)\, \mathrm{d}s}{\bar{A}} \right\} 
= \frac{1}{\bar{A}} \int_{S} A(s)\gamma(s)\, \mathrm{d}s - \frac{1}{2\bar{A}^2} \int_{S}\int_{S} A(s)A(t)\gamma(s-t)\, \mathrm{d}s\, \mathrm{d}t,$$
{which is a simple extension of} \citet[p 67-69]{Wackernagel2003}. Plugging this into~\eqref{formula_theta_BR} yields
formula~\eqref{theta_BR}.

For censored likelihood inference in Section~\ref{subsec:cens} and conditional or 
unconditional simulation described in Section~\ref{sec:simu}, the multivariate 
limit behavior of different functions is required.  We consider here the case that
is used in the application, namely that the aggregation functionals are either
spatial averages over compact regions $S_l\subset S$, $l = 1,\dots, L$, or point
evaluations at locations $s_k \in S$, $k=1,\dots,K$, i.e.,
 \begin{align}\label{ell_def}
 \ell_j(X) = \left\{\begin{array}{ll}
  \frac{1}{|{S_j}|}\int_{S_j} X(s) \mathrm{d}s, & j = 1,\dots L, \\
 X(s_{j -L}), &  j = L + 1, \dots, L+K.
 \end{array}\right.
 \end{align}
The vector $(\ell_{1}(X), \dots, \ell_{L + K}(X))$ then satisfies the 
assumptions of Theorem \ref{thm2} and its limiting exponent measure $\tilde V$ is the 
right-hand side of~\eqref{eq:conv-ell-gumbel}. This exponent measure is not normalized, 
since by Theorem~\ref{thm1}, $\tilde V(\infty, \dots, x_j,\dots, \infty) = \exp(-x_j + \log \theta_0^{\ell_j})$,
and $\log \theta_0^{\ell_j}$ is given by~\eqref{theta_BR} for $j=1,\dots,L$, and is
equal to $0$ for $j=L+1,\dots, L+K$. We therefore define the corresponding normalized 
exponent measure by 
\begin{align}
\notag V(x_1,\dots, x_{L + K}) 
 &= \Es\left\{ \max_{j=1,\ldots,L + K} \exp\left(-x_j +  \frac{ \ell_j[\{G(s) -\gamma(s)/2\} A(s)]}{\ell_j\{ A(s) \}}    
                  - \log \theta_0^{\ell_j}\right)\right\}\\
\label{eq:HRaggregated} &= \Es\left\{ \max_{j=1,\ldots,L + K} \exp\left(-x_j +  \frac{ \ell_j\{G(s) A(s)\}}{\ell_j\{ A(s) \}}    
                  - \frac1 2 \Var\left[ \frac{ \ell_j\{G(s) A(s)\}}{\ell_j\{ A(s) \}}\right]\right) \right\},            
\end{align}
where the second equality follows from~\eqref{formula_theta_BR}. Since all aggregation
functionals are either spatial averages or point evaluations, and the vector
$(Y_1,\dots, Y_{L+K})$ with $Y_j = \ell_j\{G(s) A(s)\}/\ell_j\{ A(s) \}$, $j=1,\dots,L+K$,
is multivariate Gaussian, we recognize in~\eqref{eq:HRaggregated} the exponent measure 
of a H\"usler--Reiss distribution with parameter matrix $\Gamma$ where
$\Gamma_{jk} = \Es(Y_j-Y_k)^2$, $j,k=1,\dots, L+K$.
We can separate $\Gamma$ in different blocks such that
\begin{align}\label{HR_matrix}
  \Gamma = \left(\begin{array}{cc}
    \{\Gamma_{jk}\}_{j,k} & \{\Gamma_{jq}\}_{j,q} \\ \{\Gamma_{pk}\}_{p,k} & \{\Gamma_{pq}\}_{p,q}\end{array}\right), \quad 
  \left\{ \begin{array}{l}
    j,k = 1,\dots, L, \\
    p,q = L+1,\dots, L+K.
  \end{array}\right.
\end{align}
We directly see that $\Gamma_{pq} = \gamma(s_{p-L}-s_{q-L})$ for $p,q = L+1,\dots, L+K$.
Since $\Gamma$ is symmetric, letting $\bar{A}_j = \int_{S_j} A(s)\, \mathrm{d}s$, $j=1,\dots, L$,
it suffices to compute
\begin{itemize}
  \item[(i)] for $j,k = 1,\dots L$,
   \begin{eqnarray*}
\Gamma_{jk} & = & \frac{1}{\bar{A}_j\bar{A}_k}\int_{S_j}\int_{S_k}A(s)A(t)\gamma(s - t)\, \mathrm{d}s\, \mathrm{d}t -   \frac{1}{2\bar{A}_j^2}\int_{S_j}\int_{S_j}A(s)A(t)\gamma(s -t)\, \mathrm{d}s\, \mathrm{d}t \\
 &  & - \frac{1}{2\bar{A}_k^2}\int_{S_k}\int_{S_k}A(s)A(t)\gamma(s -t)\, \mathrm{d}s\, \mathrm{d}t;
\end{eqnarray*}
 \item[(ii)] for $j=  1,\dots L$, $q = L + 1,\dots, L+K$,
   \begin{eqnarray*}
\Gamma_{jq}  & = & \frac{1}{\bar{A}_j}\int_{S_j}A(s)\gamma(s - s_{q -L})\, \mathrm{d}s - \frac{1}{2\bar{A}_j^2}\int_{S_j}\int_{S_j} A(s)A(t)\gamma(s -t)\, \mathrm{d}s\, \mathrm{d}t.
\end{eqnarray*}
\end{itemize}

In order to show (i), we note that for $s,t \in S$,
\begin{equation}\label{eq:extendedvariance}
 \Var  \left\{ \frac{A(s)G(s)}{\bar{A}_j} - \frac{A(t)G(t)}{\bar{A}_k}\right\} = \frac{A(s)^2}{\bar{A}_j^2}\gamma(s) + \frac{A(t)^2}{\bar{A}_k^2}\gamma(t) -  \frac{A(s)A(t)}{\bar{A}_j\bar{A}_k}\left\{\gamma(s) + \gamma(t) - \gamma(s - t)\right\},
\end{equation}
since $ \Es\{G(s)^2\}= \gamma(s)$ and $\Es\{G(s)G(t)\} = 1/2 \left\{ \gamma(s) + \gamma(t) - \gamma(s-t) \right\}$. 
We use the following \citep[cf.,][p 67-69]{Wackernagel2003}

\begin{align}
  \notag\Gamma_{jk} &=   \Var  \left\{\frac{1}{\bar{A}_j} \int_{S_j} A(s)G(s)\, \mathrm{d}s - \frac{1}{\bar{A}_k} \int_{S_k} A(t)G(t)\, \mathrm{d}t \right\} \\
\notag& =    \int_{S_j}\int_{S_k} \Var  \left\{ \frac{A(s)G(s) }{\bar{A}_j}-  \frac{A(t)G(t)}{\bar{A}_k}\, \mathrm{d}s\right\}\, \mathrm{d}s\, \mathrm{d}t -  \frac{1}{2}\int_{S_j}\int_{S_j} \Var  \left\{ \frac{A(s)G(s)}{\bar{A}_j} - \frac{A(t)G(t)}{\bar{A}_j}\right\}\, \mathrm{d}s\, \mathrm{d}t  \\
 \label{eq:aggVariance}   &  \qquad - \frac{1}{2}  \int_{S_k}\int_{S_k} \Var  \left\{\frac{A(s)G(s)}{\bar{A}_k} - \frac{A(t)G(t)}{\bar{A}_k}\right\}\, \mathrm{d}s\, \mathrm{d}t.
\end{align}
Using \eqref{eq:extendedvariance}, the first term in the last equation equals
\begin{align*}
\frac{|S_k|}{\bar{A}_j^2} &\int_{S_j} A(s)^2 \gamma(s)\, \mathrm{d}s + \frac{|S_j|}{\bar{A}_k^2} \int_{S_k} A(s)^2 \gamma(s)\, \mathrm{d}s - \frac{|S_k|}{\bar{A}_j} \int_{S_j} A(s) \gamma(s)\, \mathrm{d}s\\
 &- \frac{|S_k|}{\bar{A}_k} \int_{S_k} A(s) \gamma(s)\, \mathrm{d}s +  \frac{1}{\bar{A}_j\bar{A}_k} \int_{S_j}\int_{S_k} A(s)A(t) \gamma(s - t)\, \mathrm{d}s\, \mathrm{d}t.
\end{align*}
For the second term in \eqref{eq:aggVariance}, this simplifies to
\begin{align*}
\frac{2|S_j|}{\bar{A}_j^2} \int_{S_j} A(s)^2 \gamma(s)\, \mathrm{d}s - \frac{2|S_j|}{\bar{A}_j} \int_{S_j} A(s) \gamma(s)\, \mathrm{d}s + \frac{1}{\bar{A}_j^2} \int_{S_j}\int_{S_j} A(s)A(t) \gamma(s - t)\, \mathrm{d}s\, \mathrm{d}t,
\end{align*}
and analogously for the third term. Putting this together, we obtain the formula in (i).
Very similar calculations yield the result in (ii). 

With the above calculation we have shown that the $(L+K)$-dimensional vector in \eqref{ell_def}
is in the max-domain of attraction of a H\"usler--Reiss distribution with explicitly
known parameter matrix, and we can use the inference and simulation methodology described in
Sections \ref{sec:inference} and \ref{sec:simu} and in the literature.

\section{Estimation of the marginal normalizing constants $\mu_{j,n}$ and $\sigma_{j,n}$} \label{app:margins}

We present two classical approaches to estimate the marginal location and scale parameters $\mu_{j,t}$ and 
$\sigma_{j,t}$ in Equations \eqref{eq:mu} and \eqref{eq:sigma}, respectively,
for large $t$, namely the  peaks over threshold approach based on Equation 
\eqref{eq:exceedance} and the block maxima approach based on \eqref{eq:blockmaxima}.
These approaches can be used in the first step of the least squares estimation procedure
described in Section \ref{subsec:ls}.

As a first approach, the parameters can be obtained by a censored likelihood 
approach for exceedances over high thresholds based on Equation \eqref{eq:exceedance}.
Let $u_j$ be a suitably high marginal threshold, such as the empirical $(1-1/t)$-quantile of $\ell_j(X)$, and $\mathcal{I} = \{i = 1,\ldots,n: \ \ell_j(X_i) > u_j\}$.
We then consider the censored log-likelihood
$$ \log L_j^{{\rm (cens)}}(\mu_j,\sigma_j) \propto (n - | \mathcal{I}|) \cdot \left\{1 - \frac 1 t \exp\left(- \frac{u_j - \mu_j}{\sigma_j} \right)\right\} - | \mathcal{I} | \cdot \log (n\sigma_j)  
                            - \sum_{i\in\mathcal I} \frac{\ell_j(X_i) - \mu_j}{\sigma_j} $$
to obtain the estimate $(\hat \mu_{j,t}, \hat \sigma_{j,t}) =
\mathrm{argmax}_{\mu_j \in \mathbb R, \sigma_j > 0} \log L^{{\rm (cens)}}_j(\mu_j,\sigma_j)$.

As a second approach, we can estimate the parameters based on block maxima with
a sufficiently large block size $t \in \NN$, which, for simplicity, is assumed to
satisfy $n = t \cdot n_t$ for some $n_t \in \NN$. For the i.i.d.\ random variables
$M^{(j)}_{t,k}$, $k=1,\ldots,n_t$, defined by
\begin{equation*}
 M^{(j)}_{t,k} = \max_{i=(k-1)t + 1,\ldots,kt} \ell_j(X_i),
\end{equation*}
we have
\begin{equation*}
 \PP\left( M^{(j)}_{t,k} \leq x\right) = \left[\PP\left\{ \ell_j(X) \leq x\right\} \right]^t \approx  \exp\left\{ - \exp\left( - \frac{x - \mu_{j,t}}{\sigma_{j,t}}\right) \right\},
\end{equation*}
cf., Equation \eqref{eq:blockmaxima}. Thus, we obtain estimates 
$(\hat \mu_{j,t}, \hat \sigma_{j,t}) = \mathrm{argmax}_{\mu\in \mathbb R, \sigma > 0} \log L^{{\rm (BM)}}_j(\mu,\sigma)$
where
$$ \log L^{{\rm (BM)}}_j(\mu,\sigma) = - n_t \cdot \log \sigma - \sum_{k=1}^{n_t} \left\{\frac{M^{(j)}_{t,k} - \mu}{\sigma} + \exp\left( - \frac{M^{(j)}_{t,k} - \mu}{\sigma} \right) \right\}$$
is the Gumbel likelihood. 

In order to have comparable estimates $\hat \mu_{j,t}$ and $\hat \sigma_{j,t}$ 
obtained for different values $t_1$ and $t_2$, we can make use of the relationship
$ \left\{ \PP\left( M^{(j)}_{t_1,k} \leq x\right) \right\}^{n/t_1} = \left\{ \PP\left( M^{(j)}_{t_2,k} \leq x\right)\right\}^{n/t_2}$.
Then, the approximation \eqref{eq:blockmaxima} by Gumbel distributions yields
\begin{equation} \label{eq:rel-mu-sigma}
 \mu_{j,t_1} + \sigma_{j,t_1} \cdot \log(n / t_1)  \approx \mu_{j,t_1} + \sigma_{j,t_2} \cdot \log(n/t_2) \quad \text{and} \quad \sigma_{j,t_1} \approx \sigma_{j,t_2}
\end{equation}
for $t_1, t_2 \in \RR$ being both sufficiently large. Thus, estimators 
$\hat \mu_{j,t_2}$ and $\hat \sigma_{j,t_2}$ can be obtained by plugging 
$\hat \mu_{j,t_1}$ and $\hat \sigma_{j,t_1}$ into relation 
\eqref{eq:rel-mu-sigma}.

\begin{remark}
Note that, in both cases, the estimators $\hat\mu_{j,t}$ and $\hat\sigma_{j,t}$
are obtained independently for each $j = 1,\dots,L$ via a maximum 
likelihood approach. Thus, the estimated vector $((\hat \mu_{j,t})_{j=1}^L, 
(\hat \sigma_{j,t})_{j=1}^L)$ also maximizes the independent log-likelihood
functions
$$ \log L^{{\rm (cens)}}\{(\mu_j)_{j=1}^L, (\sigma_j)_{j=1}^L\} = \sum_{j=1}^L \log L^{{\rm (cens)}}_j(\mu_j, \sigma_j) $$
and
$$ \log L^{{\rm   (BM)}}\{(\mu_j)_{j=1}^L, (\sigma_j)_{j=1}^L\} = \sum_{j=1}^L \log L^{{\rm   (BM)}}_j(\mu_j, \sigma_j),$$
respectively. Such an independent likelihood could also be used to estimate
$\xi$ if unknown.
\end{remark}

 \section{Simulation study}\label{sec:simustudy}
 
In this simulation study we apply our downscaling approach to a simple model that
resembles the setup in the application in Section \ref{sec:application}.
We suppose that we observe independent data $X_1,\dots, X_n$ from a process $X$
on $S = [0,5]^2$, but only through aggregating functionals $\ell_j$, $j=1,\dots, L$, with $L = 25$, which
we will take to be spatial averages.
The observations are thus $25$-dimensional and of the form 
$$\left(\frac{1}{|S_1|}\int_{S_1}X_i(s)\, \mathrm{d}s, \dots, \frac{1}{|S_{25}|}\int_{S_{25}}X_i(s)\, \mathrm{d}s\right), \quad i=1,\dots,n,$$
where 
$S_{j} = [s^j_1,s^j_1+1] \times [s^j_2,s^j_2+1]$, with $s^j_1,s^j_2 \in \{ 0,\dots,4\}$, 
i.e., a regular grid of $1 \times 1$ squares. We consider $X$ in the Gumbel ($\xi=0$)
max-domain of attraction of a Brown--Resnick process associated  to the semi-variogram model
$$ \gamma(s,t) = \left( \frac{\|s - t\|_2}{\lambda} \right)^\alpha, \qquad \alpha = 1.5, \lambda=1. $$  
We impose a linear structure on the unknown functions $A$ and $B$ of the margins
appearing in the setting described in Section \ref{subsec:setting},
$$ \begin{array}{lll}
A(s_1,s_2) & = a_0 + a_1 \times s_1 & = 0.8 + 0.4 \times s_1 , \\
B(s_1,s_2) & = b_0 + b_2 \times s_2 & = -0.4 + 0.8 \times s_2,
\end{array} \qquad (s_1,s_2) \in [0,5]^2, 
$$  
where the parameters were chosen such that $\ell_1\{A(s)\} = 1$ and $\ell_1\{B(s)\} = 0$.

By Theorem \ref{thm2} and Example \ref{ex_BR_mult}, the vector of 
aggregated data $(\ell_{1}(X), \dots, \ell_{L }(X))$ is in the max-domain of attraction of a multivariate 
H\"usler--Reiss distribution with dependence matrix $\Gamma$ 
described in Section \ref{sec:exp_formula} and normalizing vectors $\{\mu_{j,t}\}_{j=1}^L$
and $\{\sigma_{j,t}\}_{j=1}^L$ as given in Equations \eqref{eq:mu}
and \eqref{eq:sigma}, respectively. Such a vector of aggregated data
can be simulated as follows.
 \begin{enumerate}
  \item Randomly select $j_0 \in \{1,\dots, 25\}$.
  \item Generate a univariate exponential variable $U\sim \text{Exp}(1)$.
  \item Generate a $24$-dimensional Gaussian vector $G$ with covariance matrix
        $\Sigma = (1/2)\{ \Gamma_{jj_0} + \Gamma_{kj_0} - \Gamma_{jk}\}_{j,k \neq j_0}$ and mean 
        $\mu = -\{\Gamma_{jj_0}/2 \}_{j \neq j_0}$.
  \item Set $\tilde G_{j_0} = 0$, $\tilde G_{-j_0}  = G$  
        and  $\tilde Y  = \{U + \tilde G  -  \log\|\exp(\tilde G)\|_1\} + \log 25$.
  \item Set $Y_j = \ell_j(A) \{\tilde Y_j + \log \theta_0^{\ell_j}\} + \ell_j(B)$ for $j = 1,\dots,25$.
  \item Return $Y = (Y_1, \dots, Y_{25})$.
\end{enumerate}
We simulate $n = 10^4$ samples $Y_1,\dots, Y_n$ of the random vector $Y$, which are then used to estimate the 
model parameters via the least squares and censored likelihood procedures.
Note that we could have simulated the process $X$ on a fine grid and then aggregated it
over the squares. This approach is computationally very inefficient and gives essentially
the same results as those presented in the sequel.

For the least squares procedure, we use the block maxima approach grouping the 
$10000$ replicates are into blocks of size $100$ in which we compute the maxima, 
yielding theoretically $a(t) = 1$ and $b(t) = \log(100)$ for $t=100$. We also compute
spatial means for larger squares with side length $2,3,4,5$ and then estimate the scale and location 
parameters using the univariate estimator described in Section \ref{subsec:ls}. The least squares 
fit is then performed based on the estimated scales and locations of the spatial means 
over the different squares.

For the censored likelihood method, we choose a threshold vector $u \in \RR^L$,
based on local empirical quantiles, such that the number $N_u$ of observations $Y_i$, 
$i = 1,\dots,n,$ with $\max(Y_{i1}/u_1, \dots, Y_{i 25}/u_L)  \geqslant 1$
equals $100$. In other words, we keep the $100$ highest exceedances. In this 
setting, we have $a(t) = 1$ and $b(t) = \log(10000)$ with $t=10000$. Finally, we
use these $100$ exceedances in the censored likelihood procedure described by 
Equation \eqref{eq:pplikelihood}.
  
The results in Table \ref{tab:simresults} show that all parameters can be estimated accurately. 
The censored likelihood approach, which makes use of the multivariate tail distribution, 
outperforms the least squares procedure, which relies only on marginal properties,
by about $30\%$. The advantage would be even larger if we would use more than $100$
exceedances, but we fixed this number to equal the number of block maxima.

\begin{table}[!h]
\begin{center}
\begin{tabular}{c c c c c c c c c c}
  & $a(t)$ & $a_0$ & $a_1$ & $b(t)$ & $b_0$ & $b_2$ & $\alpha$ & $\lambda$ & Mean \\ \hline
Censored LLH & $8.4$ & $1.1$ & $4.2$ & $3.0$ & $8.7$ & $8.7$ & $4.4$ & $9.4$ & $6.4$\\ 
Least squares  & $7.6$ & $2.7$ & $11.4$  & $6.6$  & $7.7$ & $7.7$ & $5.3$ & $10.2$ & $8.1$\\ 
\end{tabular}
\end{center}
\caption{Relative root mean square error (\%) for estimates based on censored 
         likelihood and least squares procedures. Inference is performed based
         on the $n=10^4$ simulated data above.} 
\label{tab:simresults}
\end{table}

\newpage

\section{Model Assessment}\label{app:assessment}

\begin{figure}[h!] 
\begin{center}
\begin{tabular}{c}
\includegraphics[width=\textwidth]{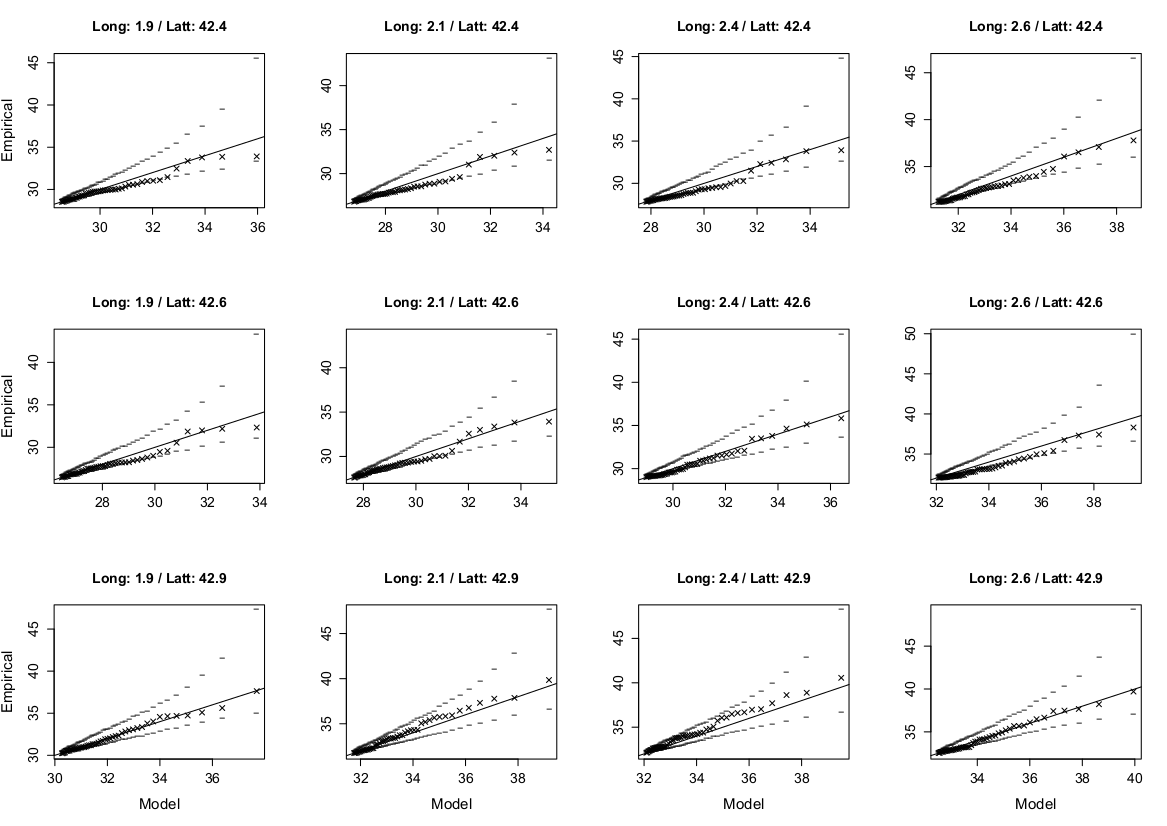}
 \end{tabular}
\end{center}
\caption{Quantile-quantile plots comparing the observations and the fitted marginal distribution
         for every grid cell. Pointwise confidence intervals are obtained by parametric bootstrap
         taking into account the uncertainty of the parameter estimates.} \label{fig:qqplot}
\end{figure}

\begin{figure}[h!] 
\centering
\begin{tabular}{ccc}
 \includegraphics[scale = 0.22]{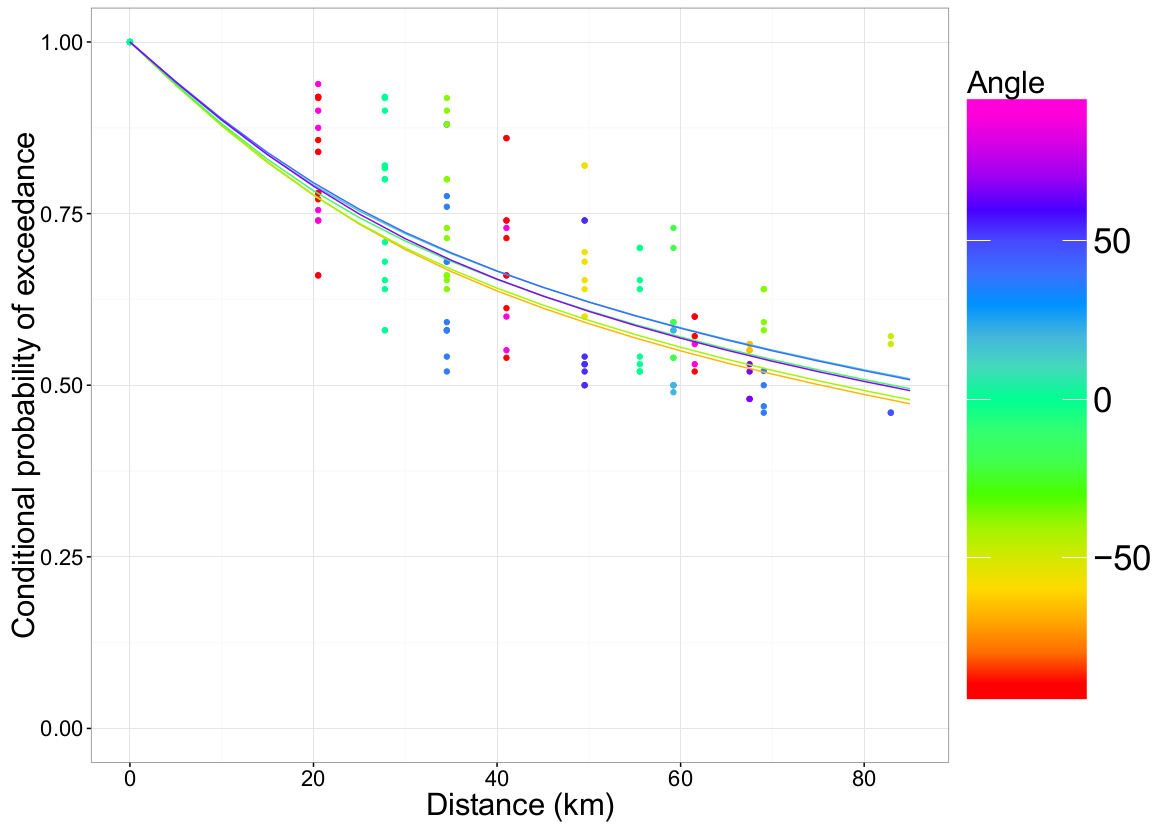}\end{tabular}
 \caption{Estimated pairwise extremogram (dots) as function of the distance (km) between the centers of the grid cells 
          and direction ($^{\circ}$). 
          The solid lines represent the theoretical extremogram for the estimated anisotropic power variogram.} \label{fig:dep}
\end{figure}

\end{document}